\documentclass[fleqn,10pt]{wlscirep}
\usepackage[utf8]{inputenc}
\usepackage[T1]{fontenc}
\usepackage{amsmath,amsthm,amssymb}
\usepackage{mathtools}
\usepackage{blkarray}
\title{Coordination and equilibrium selection in games: the role of local effects}

\author[1,2,*]{Tomasz Raducha}
\author[1]{Maxi San Miguel}
\affil[1]{Instituto de F\'isica Interdisciplinar y Sistemas Complejos IFISC (CSIC-UIB), Palma, Spain}
\affil[2]{Institute of Experimental Physics, Faculty of Physics, University of Warsaw, Warsaw, Poland}

\affil[*]{tomasz@ifisc.uib-csic.es}


\begin{abstract}
We study the role of local effects and finite size effects in
reaching coordination and in equilibrium selection in different types of two-player coordination games.
We investigate three update rules -- the replicator dynamics (RD),
the best response (BR), and the unconditional imitation (UI) -- for coordination games on random graphs. Local effects turn out to me significantly more important for the UI update rule.
For the pure coordination game with two equivalent strategies
we find a transition from a disordered state to a state of full coordination for a critical value of the network connectivity. The transition is system-size-independent for the BR and RD update rules. For the IU update rule it is system size dependent, but coordination can always be reached below the connectivity of a complete graph.
We also consider the general coordination game which covers a range
of games, such as the stag hunt. For these games there is 
a payoff-dominant strategy and a risk-dominant strategy with associated states of equilibrium coordination. We analyse equilibrium selection analytically and numerically. For the RD and
BR update rules mean-field predictions agree with simulations and the risk-dominant
strategy is evolutionary favoured independently of local effects. When players use the unconditional
imitation, however, we observe coordination in the payoff-dominant strategy.
Surprisingly, the selection of pay-off dominant equilibrium only occurs below a critical value of the network connectivity and it disappears in complete graphs.
As we show, it is a combination of local effects and update rule that
allows for coordination on the payoff-dominant strategy.
\end{abstract}
\begin{document}

\flushbottom
\maketitle

\thispagestyle{empty}

\section*{Introduction}

Voluntary individual efforts are often
crucial for the collective survival.
Cooperation might be equally important in an economic system, a small social group,
or a transportation network, increasing efficiency of personal actions.
In many man-made
systems proper regulations enforce that we end up in an optimal configuration,
i.e. that we coordinate. However, in some cases global supervision is impossible
and local willingness to cooperate does not ensure coordination.
Nevertheless, in many complex systems coordination emerges
without an external control or regulation. Bottom-up initiatives shaping our world
regularly arise without any supervision, driven purely by local interactions.
Understanding this phenomenon
is of a paramount importance in the context of global challenges that require
coordinated efforts, like pandemics \cite{bhattacharyya2011wait,brune2020evolutionary}
and climate change \cite{barfuss2018optimization,barfuss2020caring},
but also in explaining our evolution and in the emergence of language
\cite{steels1995self,dall2006nonequilibrium,selten2007emergence}.

The increase in size and complexity of social structures gives rise to the question
of how to organise them in harmony. Any solution to this problem requires
an understanding of which local processes drive human coordination and
how they can be used to reinforce the unity of action.
These questions are often studied by analysing so-called coordination games
\cite{weidenholzer2010coordination,antonioni2013coordination,mazzoli2017equilibria,antonioni2014global}.
In such framework, two or more agents are involved in a game whose outcome is
more profitable for everyone if they choose to play the same strategy.
This approach helps explaining outcomes of coordination dilemmas. The question of equilibrium selection refers to the 
underlying trade off between benefit (payoff dominance) and safety
(risk dominance) which is the essence of many organisational and economic challenges.

Our approach is based on evolutionary game theory
\cite{sigmund1999evolutionary,axelrod1981evolution,nowak1992evolutionary,nowak2006five}.
A population of players
can interact through a network of connections. An interaction consists of
playing a game specified by a payoff matrix. The strategy played by individuals
evolves over time, as they adapt to the strategies played by their neighbours
and their own payoff or fitness. The players always try to increase their payoffs,
or in other interpretation those with higher fitness spawn more replicas of themselves.
Evolutionary game theory has applications in a range
of fields, from economics to sociology to biology
\cite{friedman1998economic,newton2018evolutionary,hammerstein1994game}.
Usually, in theoretical or computational analysis, an arbitrary procedure of updating the strategy of individuals during the evolutionary game is considered. Here, we also explore the impact of the selected update rule on emergence of coordination
among agents in different games.
Local effects, however, can have a crucial influence on the coordination.
We test a wide range of connectivity
and system sizes to see which features have the largest impact on the
collective behaviour.

We study equilibrium selection under different update rules by
investigating a spectrum of coordination games played in a population of agents.
We look at their behaviour in a network of $N$
nodes, each with a degree $k$, i.e. connected to other $k$ agents in the network.
We consider a fixed structure of networks only, temporal
networks \cite{holme2012temporal,li2017fundamental}
and coevolving networks \cite{gross2008adaptive,perc2010coevolutionary}
are out of the scope of this article.  
Our numerical simulations are performed for random regular graphs
\cite{newmannetworks}, unless specified otherwise.
We analyse two-player games, i.e. only two players can be involved
in the game at a time. Such games are described by a $2 \times 2$ payoff matrix:
\begin{equation}
\begin{blockarray}{ccc}
 & $A$ & $B$  \\
\begin{block}{c(cc)}
  $A$ & R & S  \\
  $B$ & T & P  \\
\end{block}
\end{blockarray}~,
\label{eqn:matrix_most_general}
\end{equation}
where A and B are two strategies available to the players
(we only consider pure strategies).
Parameters $R$, $S$, $T$, and $P$ define the payoffs of the row player
and don't change during the evolution of the system. Note, that coordination games
are defined by $R>T$ and $P>S$. A special case of a coordination game is the stag hunt game for which $R>T>P>S$ \cite{skyrms2004stag}. In many cases we can distinguish 
a payoff-dominant (aka Pareto-optimal) and a risk-dominant strategy
\cite{harsanyi1988general}.
A payoff-dominant strategy is
the one which gives the largest absolute payoff. A risk-dominant strategy
is the one which gives the largest expected payoff,
assuming that the opponent will play either strategy
with equal probabilities. This means that the risk-dominant strategy
is the safest option in lack of information.
For the payoff matrix~(\ref{eqn:matrix_most_general}) the strategy
A is payoff-dominant if $R>P$, and the strategy B is risk-dominant
if $R-T<P-S$.
When all players coordinate on one of these strategies we 
refer to such state as a payoff-dominant or risk-dominant equilibrium.

We run numerical simulations with agents initially using a random strategy.
At the beginning of each simulation every node is assigned a strategy A or B
with equal probability. Then, before starting
with the actual algorithm each agent plays the game with every neighbour
and receives the corresponding payoff. This procedure sets up the initial payoff
for the simulation. Note, that assigning any fixed initial payoff (like zero) to every
node instead of establishing it based on the initial strategies
can lead to considerably different results.
To analyse the time evolution of the system, we update the state of nodes,
i.e. the last payoff and the used strategy, in each time step until a stationary state
or a frozen configuration is reached.
We use random asynchronous update which prevents formation of trapped oscillating states.
In every time step a node, called a focal or active node, is randomly selected
to play the game with its neighbours and update its strategy based on its
payoff and the applied update rule.
There are several update rules used in the literature on evolutionary game theory
\cite{szabo2007evolutionary}. We use three of the most common ones:
\begin{itemize}
    \item {the Replicator Dynamics (RD)} (aka replicator rule, or proportional imitation rule) -- the active node compares the payoff with a random neighbour and copies its strategy with probability $p=(\mathrm{payoff~diff.})/\phi$, if the neighbour's payoff is bigger. Normalisation $\phi$ is the largest possible payoff difference allowed by the payoff matrix and network structure and it sets the probability $p$ within $[0,1]$ range,
    \item {the myopic Best Response (BR)} -- the active node chooses the best strategy given the current strategies of the neighbours, i.e. it compares all payoffs it would obtain playing each possible strategy against the current strategies of the neighbours and chooses the strategy resulting in the largest payoff,
    \item {the Unconditional Imitation (UI)} -- the active node copies the strategy of the most successful neighbour, i.e. the one with the highest payoff, if its payoff is larger.
\end{itemize}

Another relatively popular rule is the Fermi update rule
\cite{blume1993statistical,traulsen2007pairwise}.
It is similar, however, to the replicator dynamics
in nature, but introduces an additional parameter
accounting for noise or temperature.
The replicator dynamics can be seen mostly in biological applications
\cite{schuster1983replicator,hammerstein1994game,nowak2004evolutionary},
whether to describe replication
of genes, species, or individuals. The best response update rule is usually considered
in the economics literature
\cite{kandori1993learning,young1993evolution,blume1995statistical,ellison1993learning,sandholm1998simple,buskens2008consent},
where the assumption of rational
agents is typical. The unconditional imitation
is popular in complexity and social science \cite{nowak1992evolutionary,vilone2012social,vilone2014social,lugo2015learning,gonzalez2019coordination,lugo2020local},
as it resembles social imitation. Surprisingly, all three update
rules can lead to substantially different outcomes,
therefore the evolutionary game environment
is defined by the payoff matrix as much as by the update rule
\cite{ohtsuki2006replicator,roca2009evolutionary,xia2012role,szolnoki2018dynamic,danku2018imitate,poncela2016humans}.
Consequently, it is crucial to understand the implications of each
update rule in order to be able to compare it to empirical results
and identify rules that are actually used by humans. Both experimental
and theoretical work are necessary to face this challenge.
While many experiments are being performed to uncover the mechanisms
behind human decision making when playing games
\cite{camerer2011behavioral,berninghaus2002conventions,gracia2012heterogeneous,cuesta2015reputation,frey2012equilibrium}, we address here the theoretical implications of different updating mechanisms.

We focus on the role of local effects and finite size effects
on reaching coordination 
and equilibrium selection 
under different update rules. A number of previous results, described later in detail, suggest that local effects are more important for update rules
which have an imitative nature, such as unconditional imitation.
\cite{alos2006imitation,ohtsuki2006replicator,roca2009evolutionary}.
A particularly important specific result \cite{alos2006imitation}, establishes that above a threshold degree of connectivity of agents placed on
a circle and using unconditional imitation the Pareto-optimal
equilibrium is selected. We find here that this effect
is reversed in random networks -- the Pareto-optimal equilibrium
is only selected below a threshold degree of connectivity in the network. This highlights the need of a quantitative and detailed study of local effects in equilibrium selection. 

In order to measure the level of coordination
we incorporate a parameter $\alpha \in [0,1]$,
called coordination rate,
accounting for the fraction of nodes using the strategy A. Therefore,
$\alpha = 1$ indicates a full coordination in the system with every agent using the 
strategy A, while $\alpha = 0$ indicates a full coordination on the strategy B.
To characterise the
evolution of the system we additionally
use the fraction of active connections $\rho \in [0,1]$ (interface density) and
the convergence time $\tau$. An active connection is a link between two 
nodes in different states, i.e. using different strategies. The convergence
time is the time at which the system achieves a frozen
configuration, always counted in Monte Carlo (MC) time steps.
If instead of the frozen configuration the system
stabilises in an active stationary state, the convergence time $\tau$ will reach
the maximal limit indicated in individual cases.


\section*{Previous results in coordination games}

Coordination games have been extensively studied in the past.
In particular, in two-player games the competition between
a payoff-dominant and risk-dominant equilibrium gained a lot of attention. 
When the payoff matrix is constructed in such a way that one strategy is payoff-dominant and the other is risk-dominant,  it is in principle not clear
which strategy will be favoured by the population. Therefore, a number
of authors have investigated how a stochastically  stable  equilibria
is chosen and approached by the system \cite{foster1990stochastic}.

Most of the work has been focused on the myopic best response update
rule, with some interest in imitation rules as well
\cite{weidenholzer2010coordination}. The KMR model
explored the equilibrium selection in well mixed populations
with players updating their strategies according to the best response
rule \cite{kandori1993learning}. The model is an equivalent of
interactions on a complete or fully connected graph. The unique long run equilibrium
was found to correspond to every agent playing the risk-dominant strategy. In other
words, the risk minimisation was more important than the profit maximisation
for players. The model was extended to include n-player games,
different sub-populations and asymmetric interactions with similar results
\cite{young1993evolution,youngindividual,ellison2000basins} (note that the concept of
risk-dominance can be generalised for n-player games).
In the KMR model the focal player, when deciding about its strategy update,
compares payoffs from two strategies in the current state of the population.
As later noticed, this process is in reality of imitative nature, since
effectively the player will use the strategy that gained the most in the last round.
The model was extended to include in the payoff computation
the fact that the focal node will change it's state and therefore influence
the future state of the system, but no differences were detected
\cite{sandholm1998simple}. Note, that for large networks the effect of this extension of the KMR model is negligible.

A well-mixed population is the first approximation of how social
structures can look like. In reality they are much more complex
\cite{newman2003social,wang2006structure,kumar2010structure}.
Note, that in the approximation of a well-mixed system the best response
and unconditional imitation will give the same results.
First extensions of the KMR model onto different topologies included
a \textit{circular city}, i.e. players placed in a circle interacting
only with a fixed number $k$ of nearest neighbours \cite{ellison1993learning}.
This version solved the problem of very slow convergence, which could undermine
the idea that the long run equilibrium can be ever obtained in the real world.
However, the basic outcome was the same -- coordination in the risk-dominant strategy
was preferred over coordination in the payoff-dominant one. This work was performed still assuming
the myopic best response. When using unconditional imitation
in the circular city model coordination in
the payoff-dominant strategy could be found
\cite{alos2006imitation}. Additionally, equilibrium selection in such case
depends on the degree $k$ of the network. There is a threshold value
$k^*$: if players are connected to more neighbours ($k>k^*$) the Pareto-optimal
equilibrium can be achieved.
An additional requirement for the minimal network size is imposed -- obviously for $k$ approaching $N$ we should expect results obtained before for complete graphs and those
favour risk-dominant equilibrium.
Nevertheless, this was the first clear evidence that equilibrium
selection can depend on the connectivity of the network (apart from 
the update rule). Note, that we report exactly opposite effect
of the connectivity in random graphs in our simulations.
Other attempts to go beyond a fully connected
population included random matching mixed with an imitative update rule
\cite{robson1996efficient}. Such an environment
favoured the payoff-dominant equilibrium as well.
However, in network science this would be an equivalent of a very
particular temporal network with random regular structure of degree $k=1$.

The work discussed above assumed synchronous update which can lead to traps of the dynamics
including oscillatory states and strange symmetric patterns, like in
the prisoners dilemma on lattice \cite{nowak1992evolutionary}.
For this reason those models allow for random mutations (errors in applying
the update rule) and theoretical results are computed in the limit
of the mutation rate going to zero. Another way of avoiding
the traps of synchronous update is using the asynchronous update, as we do in this paper,
which is stochastic by nature.

Equilibrium selection in coordination games has been later studied
on more complex networks and with various additional features imitating
real social interactions.
High clustering of the network proved to be important in equilibrium selection
for models with mutations and best response \cite{tomassini2010evolution}.
Other computational study for myopic best response showed
that the payoff-dominant equilibrium
may be chosen more frequently if the average degree of the network is larger and
the network is more centralised, i.e. the maximum degree is bigger
\cite{buskens2016effects}. This last study was
performed, however, for very small network sizes and other works didn't
confirm the importance of centrality \cite{tomassini2010evolution}
or reported inhibited cooperation in the stag hunt game under
unconditional imitation \cite{roca2009evolutionary}.
The case of random regular graphs has been investigated theoretically in the limit of infinite networks together with the limit of weak selection, i.e. with an update rule asymptotically
independent of the payoff
\cite{ohtsuki2006replicator}. The results showed that imitative update rules
in cooperation games, with probability of imitation proportional to fitness,
can favour Pareto-efficiency over risk dominance.
The same tendency was shown for unconditional imitation in random networks
with small degree for the Stag-Hunt game \cite{roca2009evolutionary}. 
In time dependent networks, which we do not consider in this paper, the possibility of changing costly
links has been explored and showed
that mixed equilibria that are not risk-dominant nor Pareto-optimal
are possible \cite{jackson2002formation}. Coevolution of node strategy and network topology in evolutionary
game theory has been reviewed elsewhere \cite{perc2010coevolutionary}.

Although equilibrium selection has been extensively studied
in the past there are still many open questions.
In general, coordination games were mostly studied under myopic best response
or on a complete graph. The Stag Hunt game is an exception, having been studied
with several update rules and for more complex
structures \cite{roca2009evolutionary,tomassini2010evolution}.
However, other possible configurations of payoff matrices were
investigated only theoretically with idealised assumptions, which
often are not feasible in social systems. A precise dependence
of equilibrium selection on the network's degree is also missing.
Numerical work was performed for very sparse networks, while theoretical
predictions were calculated for complete graphs or regular structures
like lattices and rings. Strict comparison of coordination games
on complex networks under RD, BR, and UI update rules is also lacking.
Those gaps are filled by our work.

Much less effort has been put in the study of pure coordination games, i.e. games in which the pay-off matrix is such that coordination in strategy A or B are two equivalent equilibria.
It was shown
that the average frequency found for each strategy is $1/2$, as expected
\cite{tomassini2010evolution}.
It is, however, unclear when this value comes from full coordination in each of the strategies in half of the realisations of the statistical sampling,
and when the system ends up in a non-coordinated state with half of the players
using one strategy and the other half the other strategy.
This question is especially important since it was predicted
that players using the BR update rule on regular structures
may end up in a disordered frozen configuration, i.e. without
reaching global coordination \cite{morris2000contagion}.
We explore the problem of coordination with two equivalent equilibria in
the following section.


\section*{Results}

\subsection*{Pure coordination game}

In this section we study the Pure Coordination Game (PCG)
(also known as doorway game, or driving game) in which $R=1$, $S=0$, $T=0$,
and $P=1$, resulting in a symmetric payoff matrix with respect to the two strategies:
\begin{equation}
\begin{blockarray}{ccc}
 & $A$ & $B$  \\
\begin{block}{c(cc)}
  $A$ & 1 & 0  \\
  $B$ & 0 & 1  \\
\end{block}
\end{blockarray}~.
\label{eqn:matrix_pure}
\end{equation}
There are  two equivalent equilibria for both players coordinating at
the strategy A or B (a third Nash equilibrium exists for players
using a mix strategy of 50\% A and 50\% B). As the absolute values of the payoff
matrix are irrelevant and the dynamics is defined by ratios between payoffs
from different strategies, the payoff matrix~(\ref{eqn:matrix_pure}) represents
all games for which the relation $R=P>S=T$ is fulfilled.

In the PCG the dilemma of choosing between safety and benefit
doesn't exist, because there is no distinction between risk-dominant and
payoff-dominant equilibrium. Both strategies yield equal payoffs when players
coordinate on them and both have the same punishment (no payoff) when players fail
to coordinate. Therefore, the PCG is the simplest framework to test when coordination
is possible and which factors influence it and how. It is in every player's
interest to use the same strategy as others. Two strategies, however, are present
in the system at the beginning of the simulation in equal amounts. From the symmetry
of the game we can expect no difference in frequency of each strategy being played,
when averaged over many realisations.
Still, the problem of when the system reaches full coordination in one of the strategies is not trivial. We address this question here.  

\begin{figure}[ht]
\centerline{
\includegraphics[scale=0.65]{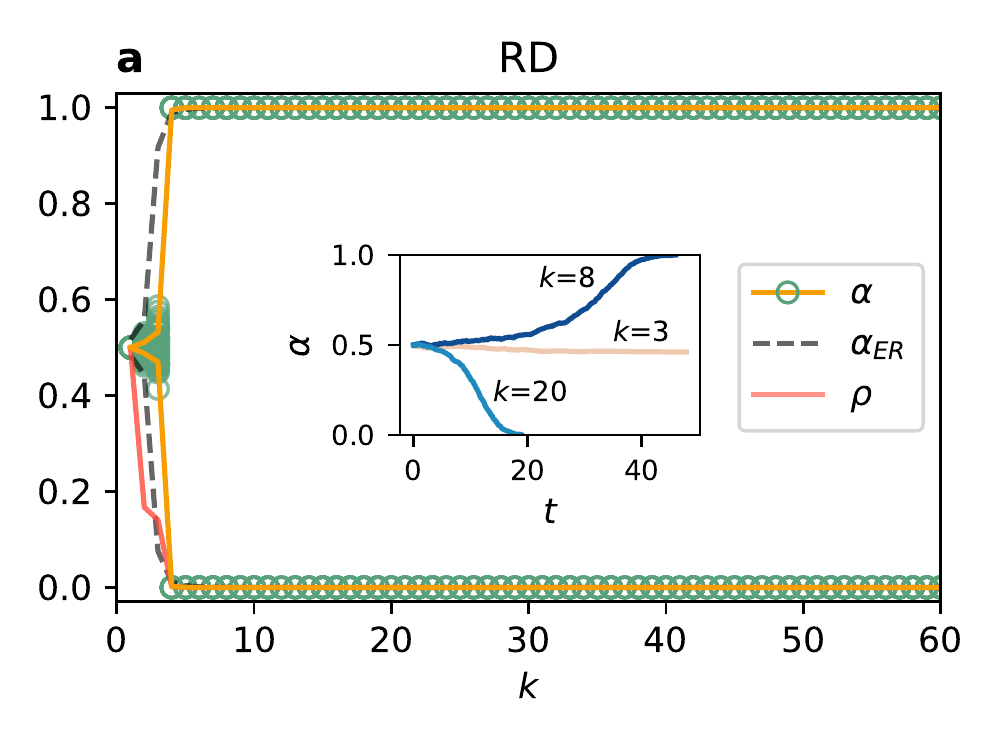}
\includegraphics[scale=0.65]{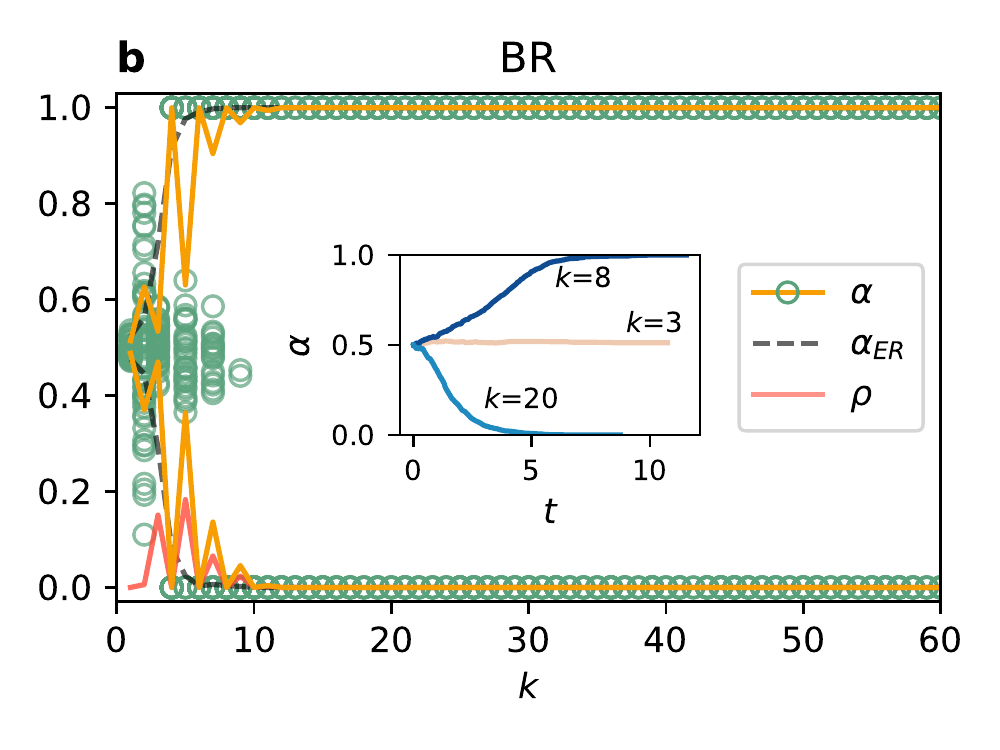}
\includegraphics[scale=0.65]{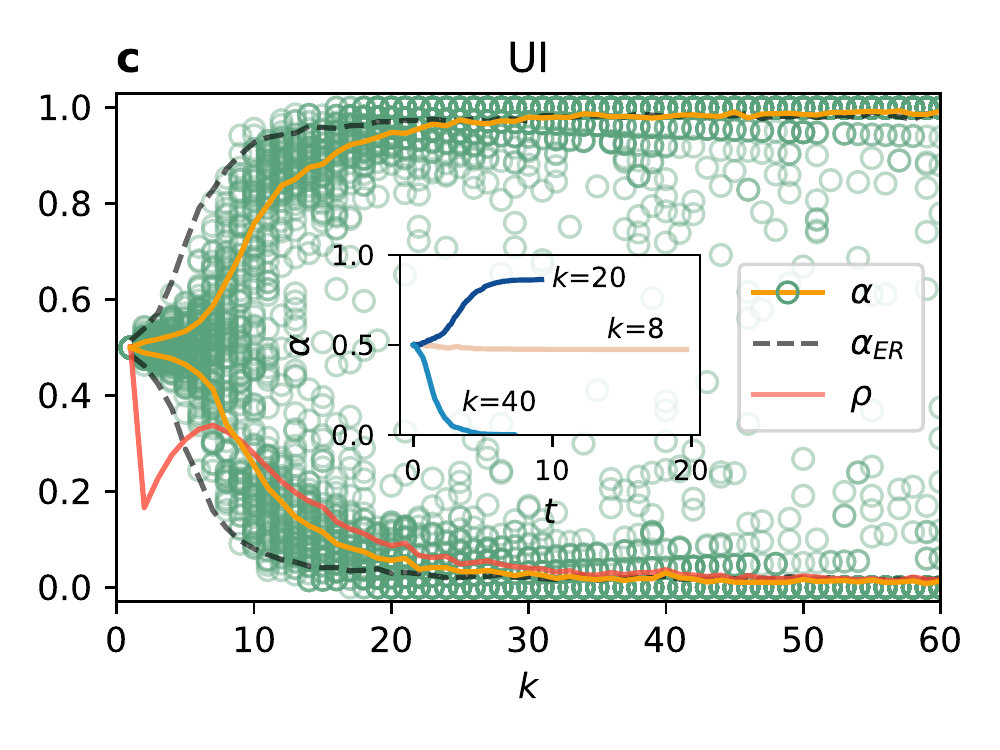}}
\caption{
Coordination rate $\alpha$ and interface density $\rho$ vs
degree $k$ of a random regular network for $N=1000$ using (a) the replicator dynamics,
(b) the best response, and (c) the unconditional imitation update rule.
Each green circle represents one of 500 realisations
for each value of the degree $k$ and 
the average value is plotted with a solid line,
separately for $\alpha>0.5$ and $\alpha \leq 0.5$. Results are compared to the ER
random network ($\alpha_{ER}$) with the same average degree. Inset plots: time evolution of $\alpha$
until obtaining a frozen configuration for 3 values of $k$.}
\label{fig:simple_all_vs_k}
\end{figure}

We present the outcome of the system's evolution in the
Figure~\ref{fig:simple_all_vs_k}.
A first thing to notice is that all plots are symmetrical with respect to
the horizontal line of $\alpha = 0.5$. It indicates that the strategies are
indeed equivalent as expected. In all cases there is a minimal
connectivity required to obtain global coordination. For RD and BR update
rules this minimum value is $k=4$, although in the case of BR the systems fails to coordinate
for small odd values of $k$ due to regular character of the graph. This
oscillating behaviour does not exist in Erd\H{o}s-R\'enyi random networks.
When nodes choose 
their strategies following the UI rule much larger values of $k$
are required to obtain full coordination. Single realisations can result
in $\alpha = 0$, or $1$ already for $k=15$. However, even for $k=60$ there 
is still a possibility of reaching a frozen uncoordinated configuration.

The important conclusion is that there is no coordination without a sufficient
level of connectivity. In order to confirm that this is not a mere artefact
of the random regular graphs we compare our results with those obtained for
Erd\H{o}s-R\'enyi (ER) random networks\cite{ER1959random,ER1960evolution}
(black dashed line in Figure~\ref{fig:simple_all_vs_k}).
The level of coordination starts to increase earlier for the three
update rules, but the general trend is the same. The only qualitative difference
can be found in the BR. The oscillating level of coordination disappears and
it doesn't matter if the degree is odd or even. This shows that different
behaviour for odd values of $k$ is due to topological traps in random regular graphs
\cite{roca2010topological}. Our results for the UI update rule are also consistent
with previous work reporting coordination for a complete graph but
failure of global coordination 
in sparse networks \cite{vilone2012social}.

\begin{figure}[ht]
\centerline{
\includegraphics[scale=0.5]{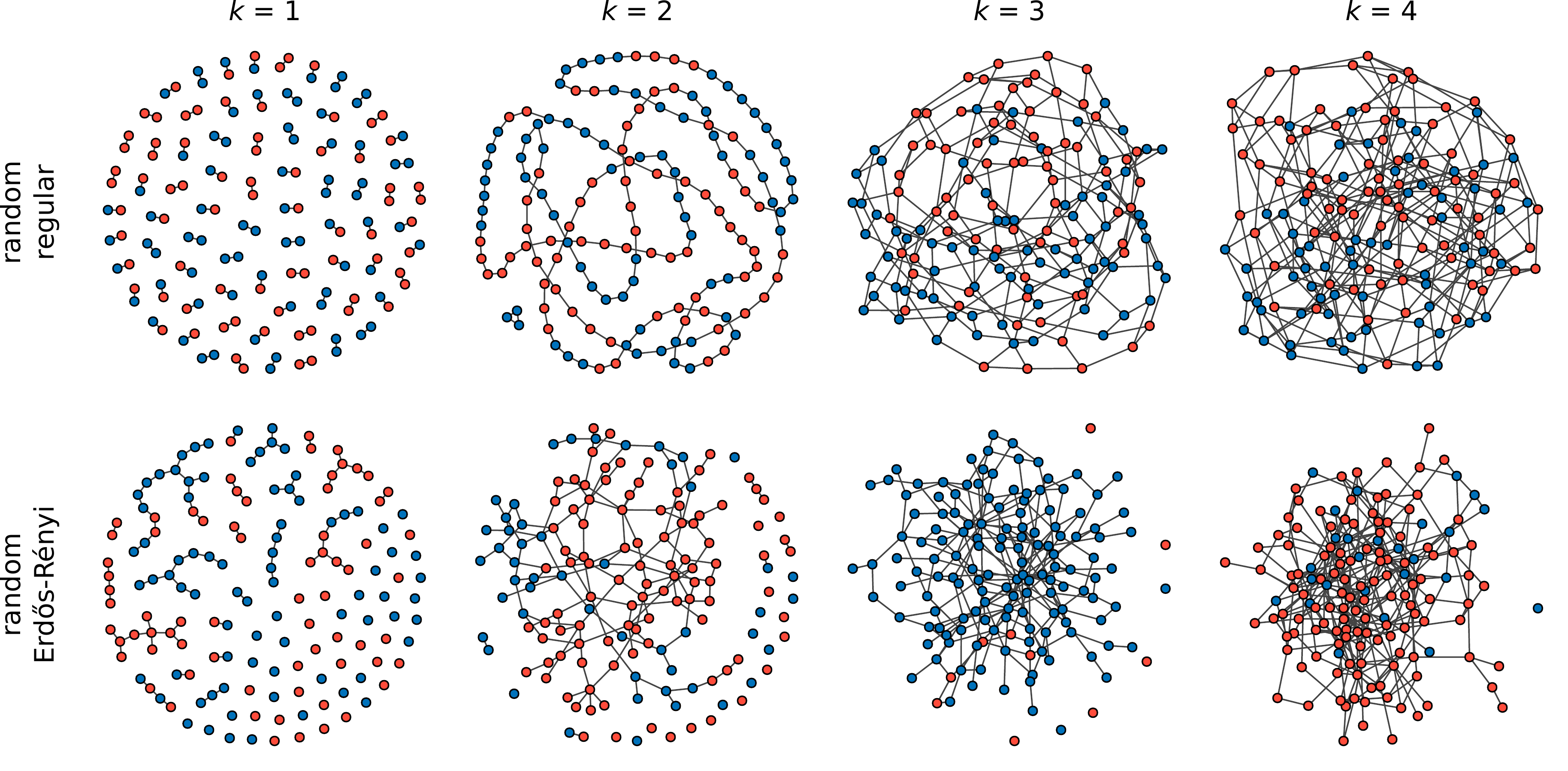}
}
\caption{Examples of frozen configuration reached under the UI update rule
for small values of the average degree $k$
in random regular networks (top row) and Erd\H{o}s-R\'enyi networks (bottom row)
with 150 nodes. Red colour indicates a player choosing the strategy A, blue colour
the strategy B. Note the topological differences between random regular and ER networks
when they are sparse. For $k=1$ a random regular graph consists of pairs of connected
nodes, while an ER network has some slightly
larger components and many loose nodes.
For $k=2$ a random regular graph is a chain (sometimes 2 to 4 separate chains),
while an ER network has one
large component and many disconnected nodes. For $k=3$ and $k=4$
a random regular graph is always composed of one component, while an ER network
has still a few disconnected nodes.}
\label{fig:networks}
\end{figure}

Since agents using the RD and BR update rule do not achieve coordination for small
values of degree, one might suspect that the network is just not sufficiently
connected for these values of the degree, i.e. there are separate components.
This is only partially true. In Figure~\ref{fig:networks}
we can see the structures generated by random regular graph and
by ER random graph algorithms. Indeed, for $k=1$ and $2$ the topology is trivial
and a large (infinite for $k=1$) average path length\cite{newmannetworks}
can be the underlying feature stopping
the system to reach coordination. For $k=3$, however, the network is well connected
with one giant component and the system still does not reach the global
coordination when using RD or BR. For the UI update rule coordination arrives even for larger values of $k$.
Looking at the strategies used by players in Figure~\ref{fig:networks}
we can see how frozen configuration without coordination can be achieved.
There are various types of topological traps where nodes with different strategies are
connected, but none of them is willing to change the strategy in the given update rule.

We next consider the question of how the two strategies are distributed in the situations in which full coordination is not reached. 
Looking at the inset plots of
Figure~\ref{fig:simple_all_vs_k} we can see that there are barely any
successful strategy updates in such scenario and the value of $\alpha$ remains
close to $0.5$ until arriving at a frozen state for $k=3$ ($k=8$ for UI).
This suggests that there is
not enough time, in the sense of the number of updates, to cluster the
different strategies in the network. Therefore, one might expect that they are well mixed
as at the end of each simulation. 
However, an analysis of the density of active links in the final state of the dynamics, also presented in
Figure~\ref{fig:simple_all_vs_k}, shows a slightly more complex behaviour.
When the two strategies are randomly distributed (i.e. well mixed) in
a network, the interface density takes the value $\rho=0.5$. When the two strategies are spatially clustered in the network
there are only few links connecting them and therefore
the interface density takes small values. Looking at the dependence of $\rho$ on $k$, we find that 
for the replicator dynamics the active link density starts at $0.5$ for
$k=1$, then drops below $0.2$ for $k=2$ and $3$ indicating good clustering
between strategies, to fall to zero for $k=4$ where full coordination is already
obtained. When using the best response update rule the situation is quite
different. For $k=1$ there are no active links, $\rho=0$, and hardly any for
$k=2$. There is a slight increase of the active link density for $k=3$, to drop
to zero again for $k=4$ due to full coordination. Because of the oscillatory level of
coordination there are still active links for odd values of $k<10$, but $\rho$
is always smaller than $0.2$. In the case of the unconditional imitation we again
start at $\rho=0.5$ for $k=1$, before it drops below $0.2$ for $k=2$. Subsequently,
the active link density grows to obtain its maximum value for $k=7$ and starts decreasing
towards zero. These differences in behaviour can be better understood when
studying the actual topology of regular graphs for small values of the degree.
In Figure~\ref{fig:networks} we present frozen configurations for the UI update rule in networks
with $k=1,2,3$, and $4$. For the smallest degree, $k=1$, the network consists of
connected pairs of nodes. Whatever strategies are initially assigned to those pairs
they will not change when using RD or UI. In both cases -- two nodes using the same
strategy and two nodes using different strategies -- both nodes in a pair receive
the same payoff, therefore no imitation can happen. Hence, the active link density is
$\rho=0.5$ for RD and UI.  On the other hand, the BR update rule will cause every
pair to coordinate, as this is the best possibility for any node, and therefore
$\rho=0$. The case of $k=2$ still results in a quite particular structure
-- it is a chain of
nodes (or a 1D lattice). For every update rule the shortest possible cluster of
one strategy must contain at least two nodes. The RD and UI separate different strategies
relatively well obtaining $\rho<0.2$, however still worse than the BR. For the latter
there are almost no active links, i.e. two strategies are perfectly clustered in two
(or only few) clusters. From $k=3$ onwards random regular graphs form well connected
networks with one giant component and small average path length. This is the largest value of $k$
for which the RD and BR update  rules do not lead to full coordination. Given the value of the active
link density $\rho<0.2$ we can say they both cluster strategies relatively well.
For the UI update rule coordination still doesn't exist for $k=3$, and
the number of active links increases with growing degree. This is to be expected --
the more links in the network the more connections between nodes playing different
strategies. The level of $\rho$ starts to drop only when the coordination begins
to settle and we observe that the strategies are well mixed in the network before this
point.

\begin{figure}[ht]
\centerline{
    \includegraphics[scale=0.53]{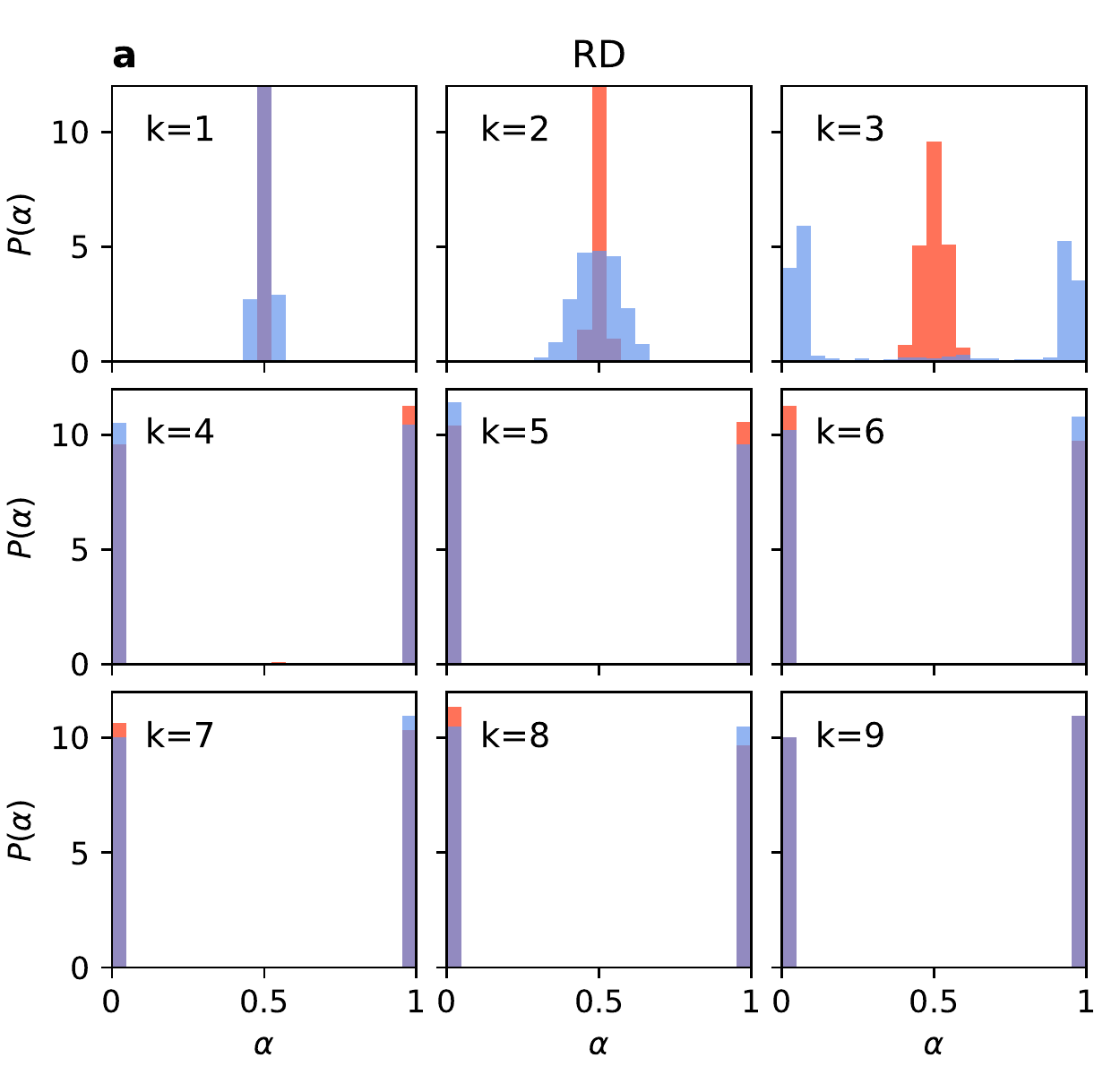}
    \includegraphics[scale=0.53]{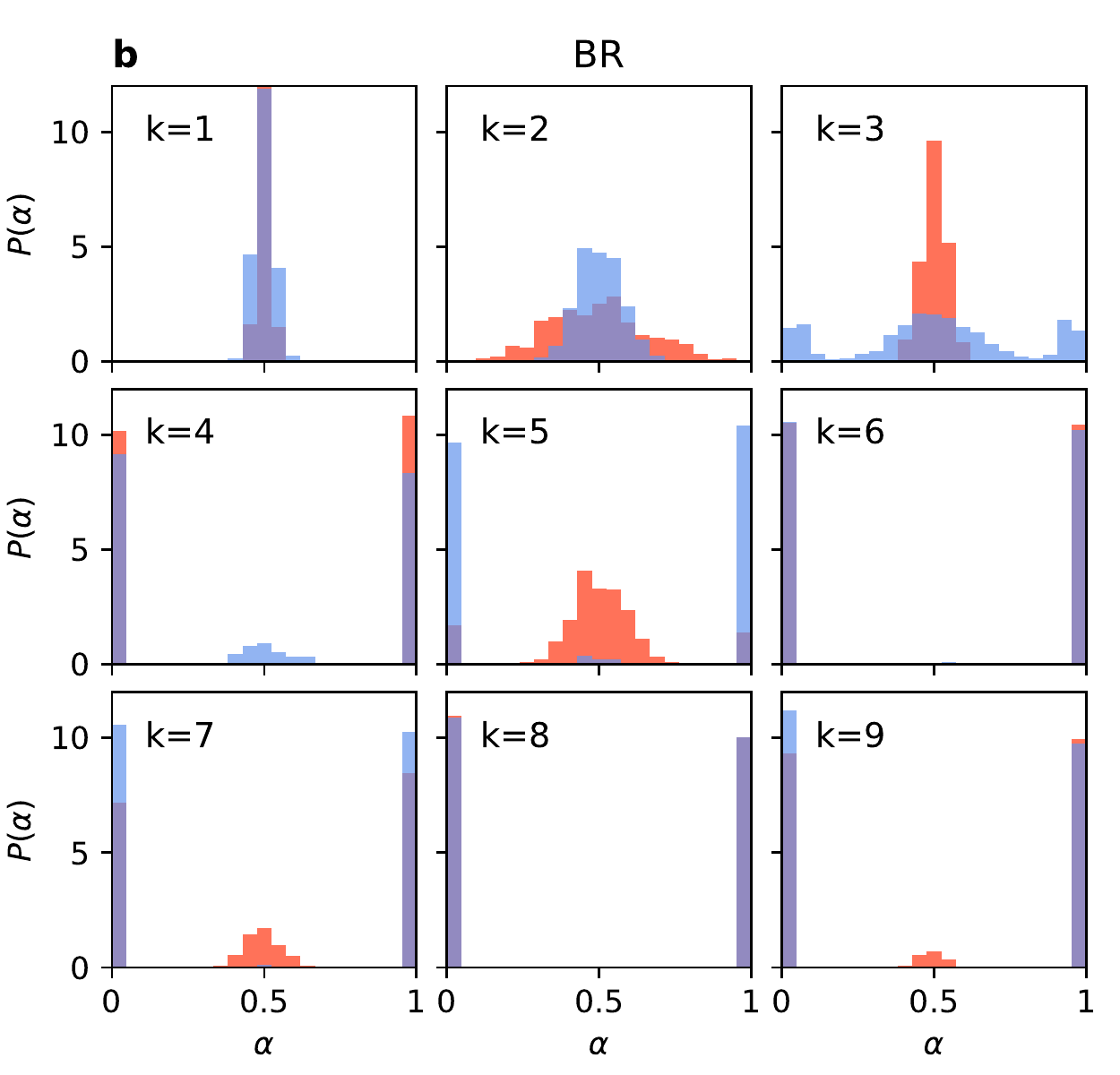}
    \includegraphics[scale=0.53]{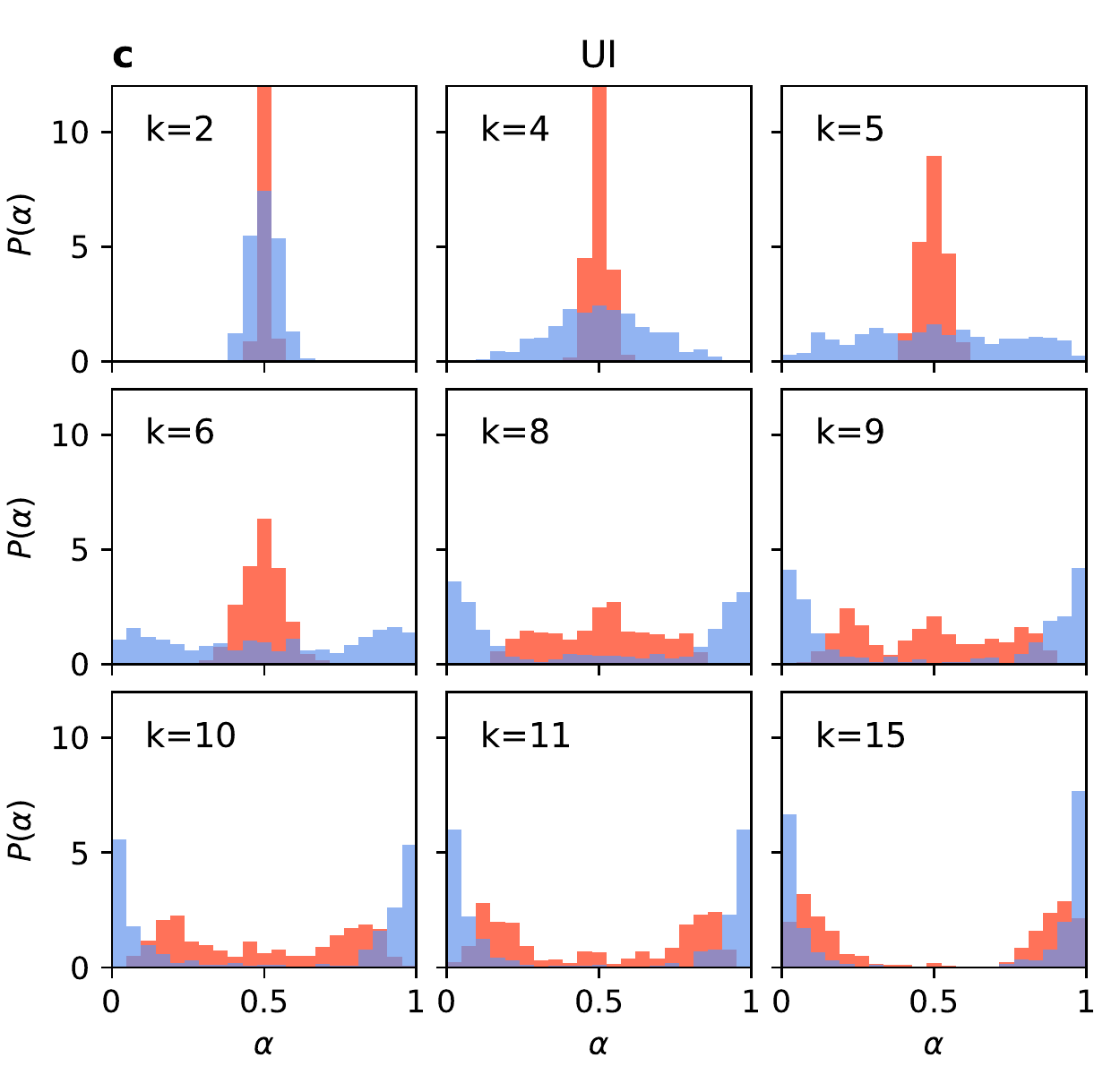}
    }
    \caption{
Distribution of the coordination rate $\alpha$ for different
values of the average degree $k$ and $N=10^3$ for (a) the replicator dynamics,
(b) the best response, and (c) the unconditional imitation update rule.
Results for random regular networks are presented in red, and equivalent results
for random ER networks in blue. Histograms are constructed from a sample of 500
realisations. Note how the distribution changes from unimodal,
via trimodal, into bimodal for the UI update rule.}
\label{fig:simple_histograms}
\end{figure}

\subsubsection*{Coordination transition}

One way of investigating the transition between frozen disorder and global coordination
is to look at the probability distribution of the coordination rate $\alpha$. If the
distribution is unimodal and centred in the middle at the value of $\alpha=0.5$,
we have a disordered state with roughly equal numbers of players using the strategy
A and the strategy B. If the distribution is bimodal with two peaks at the boundary
values of $\alpha=0$ or $1$, all players use the same strategy A or B and
coordination was reached. We can see a transition from the first scenario
into the second one for all update rules in Figure~\ref{fig:simple_histograms}. The transition point can be identified by the lowest degree for which the distribution
is not unimodal (and centred in the middle).
The specifics of these transitions, however, differ from one rule to another.
For the RD the distribution is unimodal for $k=1,2,3$, becomes trimodal
for $k = 4$ (in ER random networks for $k = 3$), and later bimodal for $k \ge 5$.
Therefore, the transition point can be defined as the threshold value of the degree $k_c^{RD}=4$. When players
use the BR update rule, the distribution becomes bimodal at $k=4$, but is trimodal
for $k=5,7,9,11$, i.e. small odd degree values. No such effect of odd degree exists
if the network is an ER random network, but also here a trimodal distribution
is obtained up to $k=9$. The transition point for the BR is the same as in the RD,
$k_c^{BR}=4$, but in the RD beyond this point coordination is always reached,
while for the BR there is still a possibility of stopping at a frozen discorded
configuration up to $k=11$. While the behavior of the probability distribution
of $\alpha$ for the RD and BR update rules is a signature of a first order transition,
the transition is less abrupt for the UI update
rule. Here the distribution of the coordination rate $\alpha$ is unimodal up
to $k=8$, although its variance increases with growing degree of the network.
At $k_c^{UI}=9$ the distribution becomes trimodal
($k=6$ for ER random networks),
but the side maxima are placed
far from the coordinated state -- rougly at $\alpha = 0.2$ and $0.8$.
The trimodal distribution is present up to $k=15$ and the side peaks keep
shifting towards boundary values. For $k>15$ the distribution is bimodal,
but the peaks are much wider than in other update rules. Additionally,
the distribution is not zero between them. This means that the
simulation sometimes freezes at a disordered configuration
or close to the global coordination,
but with a group of agents playing the opposite strategy.

\begin{figure}[ht]
\centerline{
    \includegraphics[scale=0.65]{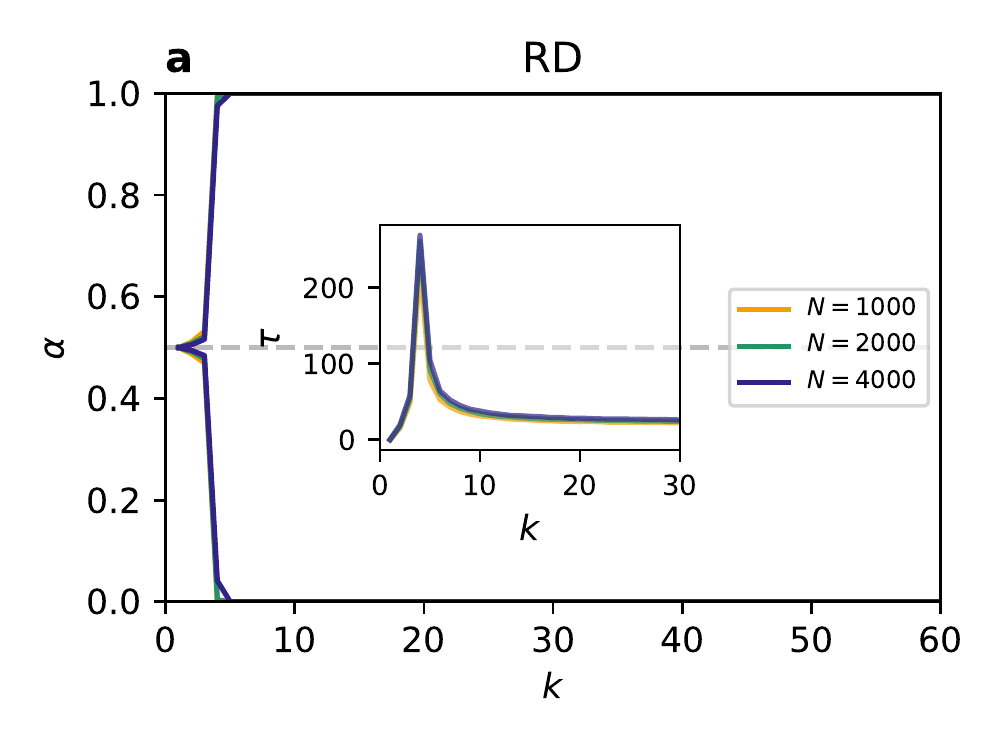}
    \includegraphics[scale=0.65]{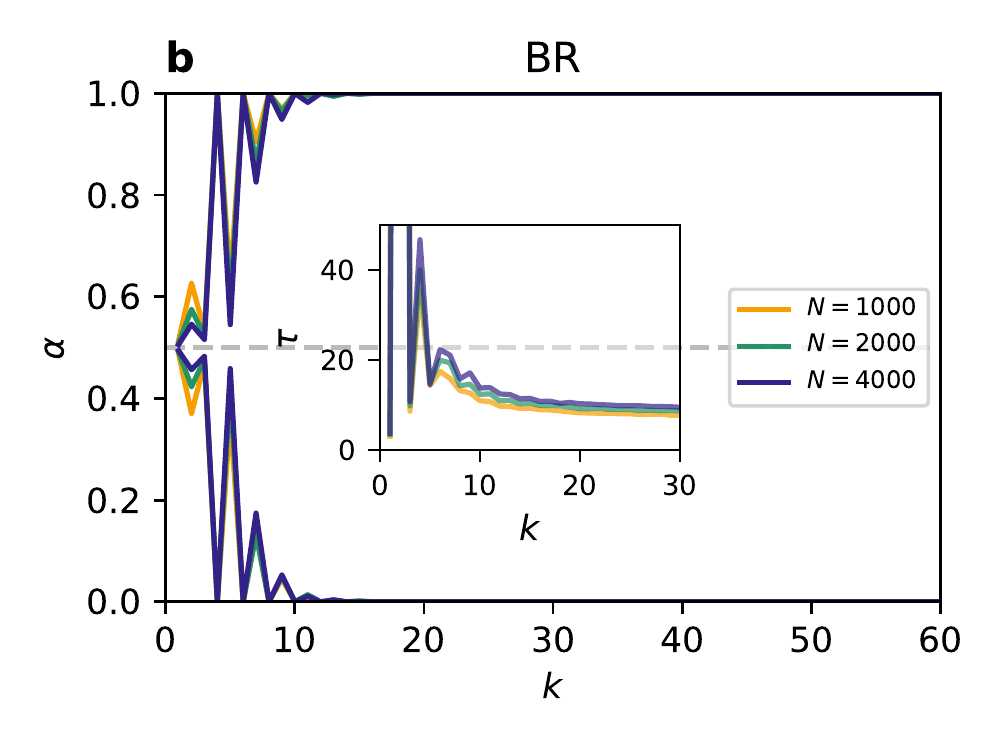}
    \includegraphics[scale=0.65]{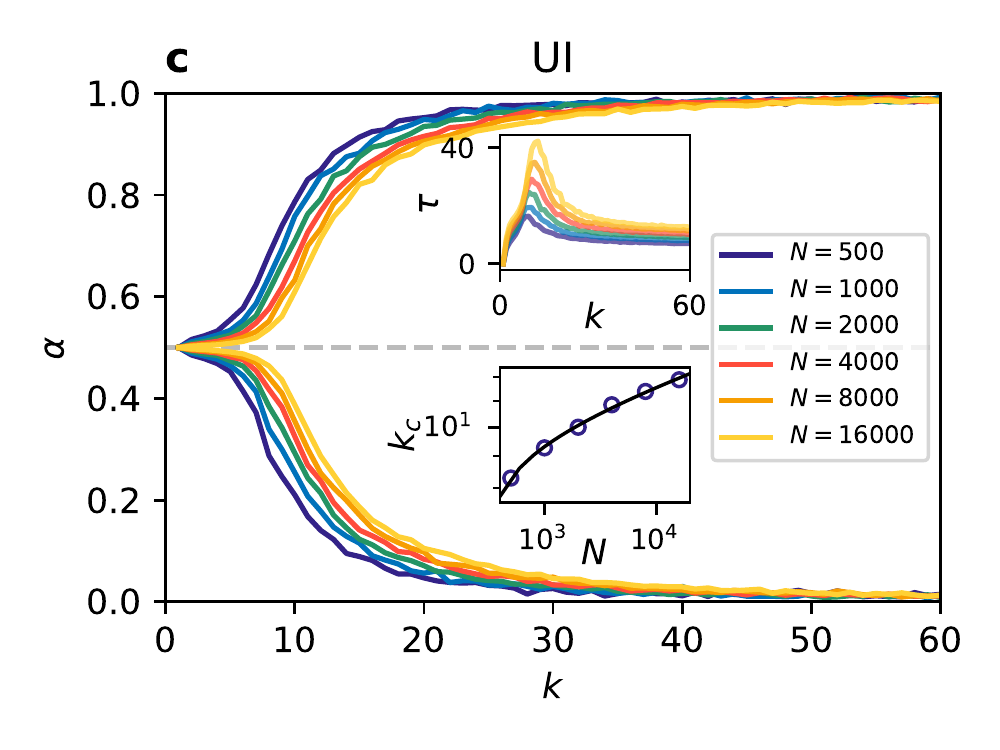}
    }
\caption{
Scaling of the average coordination rate $\alpha$ vs degree of the network $k$
for different network sizes using (a) the replicator dynamics,
(b) the best response, and (c) the unconditional imitation update rule.
The average value is computed from 500 realisations,
separately for $\alpha>0.5$ and $\alpha \leq 0.5$.
Inset plots: scaling of the convergence time $\tau$ vs $k$; additionally
in the panel (c)
scaling of the transition point $k_c$ with the system size $N$
with a logarithmic function fit in black.
Note, that the behaviour of the system does not depend on $N$
for RD and BR, but the transition into coordination shifts towards bigger $k$ with
growing network for UI.}
\label{fig:simple_scaling}
\end{figure}

An important question is what role do the finite size effects play in our results.
A larger population could need higher or lower 
connectivity to obtain the same level of coordination. The answer is given 
in Figure~\ref{fig:simple_scaling}. From panels (a) and (b) we observe that
when players use the RD or BR update rule the resulting level of coordination
for a given value of degree is the same for any system size. In other words,
even for very large populations we obtain full coordination already
for $k=4$. Interestingly, the drop in the coordination rate $\alpha$ for odd
values of $k$ visible in the BR is also the same for different system sizes.
This again suggests that it is an effect of topological traps in regular graphs
and not of finite size.

Finite size effects turn out to be quite different when players update
their strategies according to the UI update rule. From 
Figure~\ref{fig:simple_scaling} we observe that the level of coordination decreases when
the population grows at a given fixed connectivity. Or equivalently, the transition
from frozen disorder into full coordination is shifted towards higher values of
the degree $k$. A proper way of identifying the transition point
is by looking at the maximum of the convergence time. In the inset plots of
Figure~\ref{fig:simple_scaling} we see that, when increasing system size, the value of the degree for which the convergence time becomes maximum shift to larger values for the UI update rule,
whereas it stays at the same value of $k$ for the RD and BR update rules.
Additionally, the maximum number of the Monte Carlo time steps necessary to reach
a frozen configuration grows with the system size in the case of UI, but
stays the same for other update rules. The transition point $k_c$ defined
by the maximum convergence time $\tau_{max}$ is equal $k_c^{RD}=4$
for the replicator dynamics,
$k_c^{BR}=4$ for the best response, and $k_c^{UI}=9$ for the unconditional
imitation for $N=1000$ nodes. Note, that for the BR the convergence time
is higher for $k=2$, but we don't take it into account due to the trivial
topology. Those threshold values of $k$ coincide with the ones discussed above in terms of the changes in the $\alpha$ probability distribution. 

The scaling behaviour under the UI update rule raises a question
about the minimal connectivity necessary
to observe coordination in the thermodynamic limit .
In Figure~\ref{fig:simple_scaling}~(c) in the bottom inset plot
we present the dependence of the transition point $k_c^{UI}$ on
the system size $N$. Two functions can be fitted
to these data points -- logarithmic function ($R^2=0.997$) and power law
($R^2=0.973$). In the latter case
the transition depends on the number of nodes as $k_c^{UI} \sim N^{0.1}$.
For both functions the 
minimum degree required to obtain coordination goes to infinity in the thermodynamic
limit. However, $ k_c^{UI}/N  \to 0$ when $N \to \infty$ so that full coordination is achieved in the thermodynamic limit for connectivity
much lower than in a complete graph.

Our analysis of the simple PCG already uncovers
significant differences among the update rules, including different finite size effects.
To achieve global coordination in the population the
network must have a minimal connectivity. In random regular graphs this minimum
is higher than in ER random networks. Looking at the update rules, players using
the RD achieve coordination for the lowest connectivity, or in other words
the drive towards global coordination is the strongest in this update rule.
When using the BR
a slightly higher values of the degree are required for the same level of
coordination to appear. Moreover, in random regular graphs 
the minimal connectivity for the BR update rule is different for odd and even values of the degree.
For the UI update rule coordination requires  much
higher values of the degree $k$ and often freezes just before obtaining
full coordination (at $\alpha$ close but not equal $0$ or $1$). System size scaling indicates that even higher connectivity is required for coordination
to happen in larger networks, while the transition to coordination for the RD and the BR update rules is size-independent.
Interpreting the results one also has to bear in mind that cases of $k=1,2$
produce very particular topologies with large average path lengths,
which hinder the coordination. However, for $k \ge 3$ the network is well
connected and in principle there is no structural reason why coordination
should not emerge. Nevertheless, in many cases it doesn't.


\subsection*{General coordination game}

\begin{figure}[ht]
\centering
\includegraphics[scale=0.63]{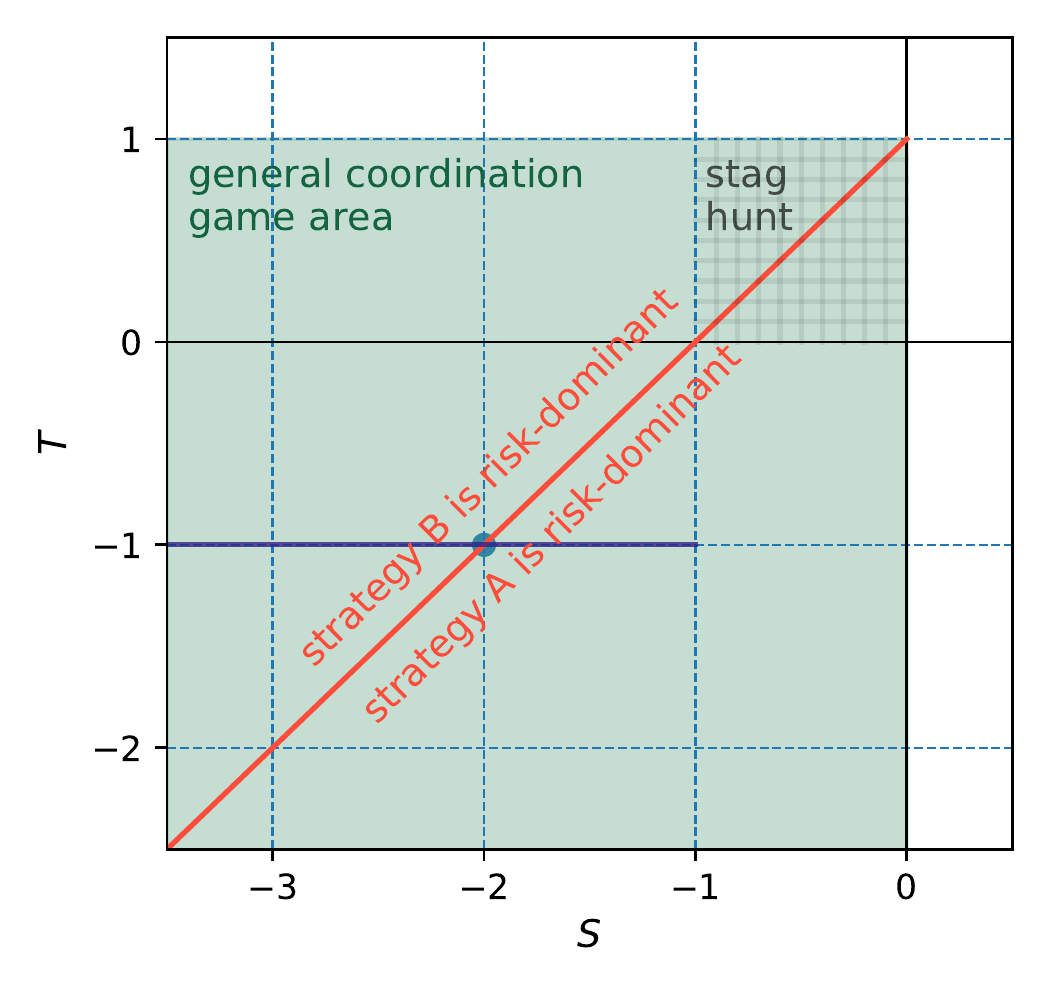}
\caption{
Parameter space of the general coordination game
described by the payoff matrix~(\ref{eqn:matrix_GCG}),
depending on parameters $S$ and $T$. The green area shows the region
of the general coordination game ($S<0$, $T<1$).
The red line represents the line in parameter space 
at which the risk-dominant strategy changes from A to B (when increasing $T$
or decreasing $S$).
The purple line represents the parametrisation from the payoff
matrix~(\ref{eqn:CCG_matrix}) for $b>0$.
}
\label{fig:diagram_games}
\end{figure}

After considering the role of local and finite size effects in the simplest
case of two equivalent equilibria for coordination, we address in this
section the question of equilibrium selection for nonequivalent
states of global coordination. We therefore consider
the payoff matrix~(\ref{eqn:matrix_most_general}) where $R \neq P$.
Without loss of generality we can assume that $R>P$
(otherwise we can rename the strategies and shuffle the columns and rows).
What defines the outcome of
a game are the \textit{greater than} and \textit{smaller than} relations
among the payoffs. Therefore we can add/subtract any value from all payoffs,
or multiply them by a factor grater than zero, without changing the game. Thus,
the payoff matrix~(\ref{eqn:matrix_most_general}) can be rewritten as:
\begin{equation}
\begin{blockarray}{ccc}
 & $A$ & $B$  \\
\begin{block}{c(cc)}
  $A$ & 1 & \frac{S-P}{R-P}  \\
  $B$ & \frac{T-P}{R-P} & 0  \\
\end{block}
\end{blockarray}~,
\end{equation}
which, after substituting $S'=\frac{S-P}{R-P}$ and $T'=\frac{T-P}{R-P}$,
is equivalent to the matrix:
\begin{equation}
\label{eqn:matrix_GCG}
\begin{blockarray}{ccc}
 & $A$ & $B$  \\
\begin{block}{c(cc)}
  $A$ & 1 & S'  \\
  $B$ & T' & 0  \\
\end{block}
\end{blockarray}~~
\xrightarrow[\text{~apostrophes~}]{\text{skipping}}~
\begin{blockarray}{ccc}
 & $A$ & $B$  \\
\begin{block}{c(cc)}
  $A$ & 1 & S  \\
  $B$ & T & 0  \\
\end{block}
\end{blockarray}~.
\end{equation}
From now on we omit the apostrophes and simply refer
to parameters $S$ and $T$.
This payoff matrix can represent many games, including e.g.
the prisoner's dilemma \cite{nowak1992evolutionary,roca2009evolutionary}
(for $T>1$ and $S<0$). 
We restrict our analysis to coordination games which correspond to $S<0$ and $T<1$ \cite{goyal2005network,buskens2016effects}. In these games there are pure
Nash equilibria at coordinated states,
i.e. for both players using the same strategy.  We call
games that fit in this range of parameters by one name -- the General Coordination Game
(GCG). In fact, it covers any two-player
symmetrical coordination game with different payoffs at coordinated states.
For example, the popular game of stag hunt \cite{skyrms2001stag,skyrms2004stag}
is a special case of the GCG
for $0<T<1$ and $-1<S<0$ (the condition $S>-1$ is not always imposed, however
conventionally the stag hunt is investigated only in the indicated square area).
For $T=-1$ we obtain a game which
in a different parametrisation is often studied in literature
\cite{eshel1998altruists,lugo2015learning}.
We shall discuss it in more detail further in the text.
Figure~\ref{fig:diagram_games} illustrates the plane of parameters of the GCG.

\begin{figure}[ht]
\centerline{
\includegraphics[scale=0.65]{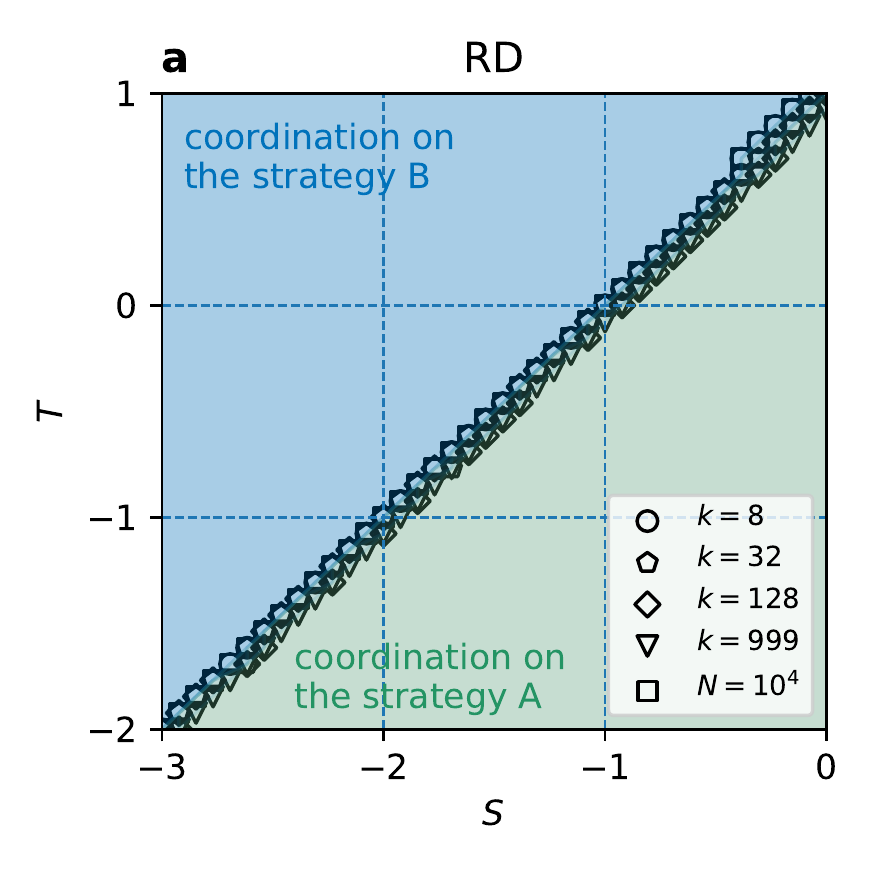}
\includegraphics[scale=0.65]{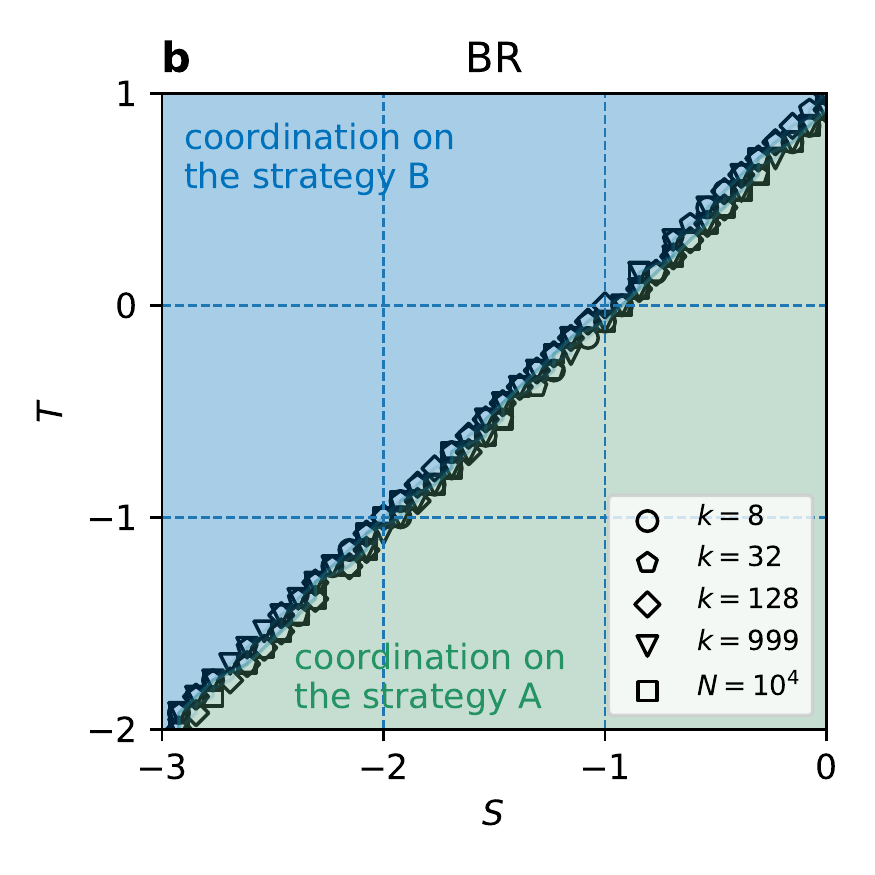}
\includegraphics[scale=0.65]{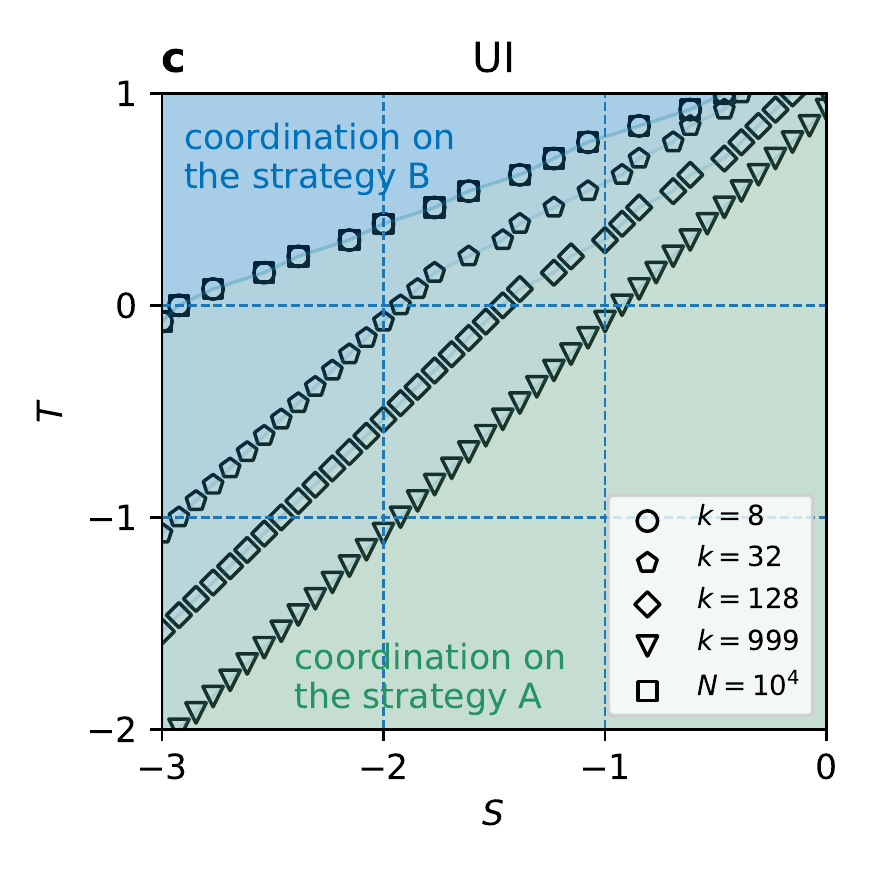}
}
\caption{Phase diagram of the general coordination game for (a) the replicator dynamics,
(b) the best response, and (c) the unconditional imitation update rule.
The green area indicates
coordination on the strategy A ($\alpha=1$) and the blue area on the strategy B
($\alpha=0$).
Note, that in the panel (c) colors blend because of the shift of transition,
but for every individual case we obtain coordination below and above 
the transition line.
The transition lines are plotted for $N=10^3$ and different values of $k$,
and for $N=10^4$ with $k=8$. Transition lines are obtained from 100 realisations,
see Supplementary Information for the full diagrams.}
\label{fig:GCG_diagram_k}
\end{figure}

An important feature of the GCG is that is possesses
a payoff-dominant (aka Pareto-optimal) equilibrium and a risk-dominant equilibrium.
The strategy A is Pareto-optimal -- it provides
the largest possible payoff for both players if they coordinate on it. To
establish which strategy is risk-dominant we need to compute the average payoffs
of strategies A and B,
$\Pi_{\mathrm{A}}$ and $\Pi_{\mathrm{B}}$, assuming random and uniform strategy choice in the population.
For both strategies having probability $1/2$ of being played,
from the payoff matrix~(\ref{eqn:matrix_GCG})
we obtain $\Pi_{\mathrm{A}} = \frac{S+1}{2}$ and $\Pi_{\mathrm{B}} = \frac{T}{2}$. Therefore, the strategy
A will be risk dominant for $T<S+1$ and the strategy B will be risk dominant
for $T>S+1$. This calculation provides a theoretical transition line $T=S+1$
between two phases, depending which of the strategies A or B 
is the risk-dominant strategy (see Figure~\ref{fig:diagram_games}).

Whether the risk dominance is a sufficient condition for a strategy
to prevail in an evolutionary setup is to be examined.
It is intuitively clear that when one strategy is both risk-dominant
and payoff-dominant at the same time it should be evolutionary favoured.
Therefore, we expect to see coordination on the strategy A for $T<S+1$.
The question is what happens when there is
a choice between a risk-dominant strategy and a payoff-dominant one. 
We explore in the following paragraphs the effect of update rule, local effects and finite size in equilibrium selection. It should be noted that we do not consider very sparse networks since from the analysis of the PCG we already know that coordination is not achieved for low enough values of the degree.

We present phase diagrams obtained from our numerical simulations for the three update rules
in Figure~\ref{fig:GCG_diagram_k}. For the replicator dynamics and
the best response update rules there is a transition in the equilibrium selection at the line $T=S+1$. For $T<S+1$ the strategy A, which is there both payoff-dominant and risk-dominant, is selected, but for $T>S+1$ the risk-dominant strategy B is selected. Importantly, this transition line from A to B selection is independent
of the degree or size of the network. These findings are also consistent
with analytical calculations based on the replicator equation for the RD
and mean field approach for the BR (see Methods for details).
Interestingly, when players use the unconditional imitation the transition
is shifted towards larger values of $T$ (and smaller $S$) for 
sparser networks. The transition line moves to $T=S+1$  with
growing degree of the network and finally reaches this line for a complete graph.
Note, that for the stag hunt area and small values of $k$ our results are
consistent with Roca et al \cite{roca2009evolutionary}.

The important consequence of local effects is that players can still coordinate on the Pareto-optimal
strategy A even when this strategy is far from being risk-dominant. However,
this only happens for the UI update rule and small enough connectivity in the population.
This is in a sense opposite to the results in the PCG. There, the optimal
outcome, i.e. any coordination, could be achieved only above a threshold value of the degree.
Here, for a given range of the parameters $S$ and $T$ the optimal outcome,
i.e. coordination on the Pareto-optimal strategy, can be achieved only
for networks with small enough connectivity. It is important to note that
a previous most explicit result on the importance of local effects 
was given for rings (circular city) \cite{alos2006imitation}.
The behaviour there was exactly
opposite -- a larger connectivity was required to coordinate on the Pareto-optimal
strategy. Including more complex structure, as we show, reverses this relation.
One might imagine that our results could change when analysing networks with the degree
smaller than 8. In such sparse networks, that we do not show here, we observe
coordination on the Pareto-optimal strategy or no coordination.

\begin{figure}[ht!]
\centerline{
\includegraphics[scale=0.6]{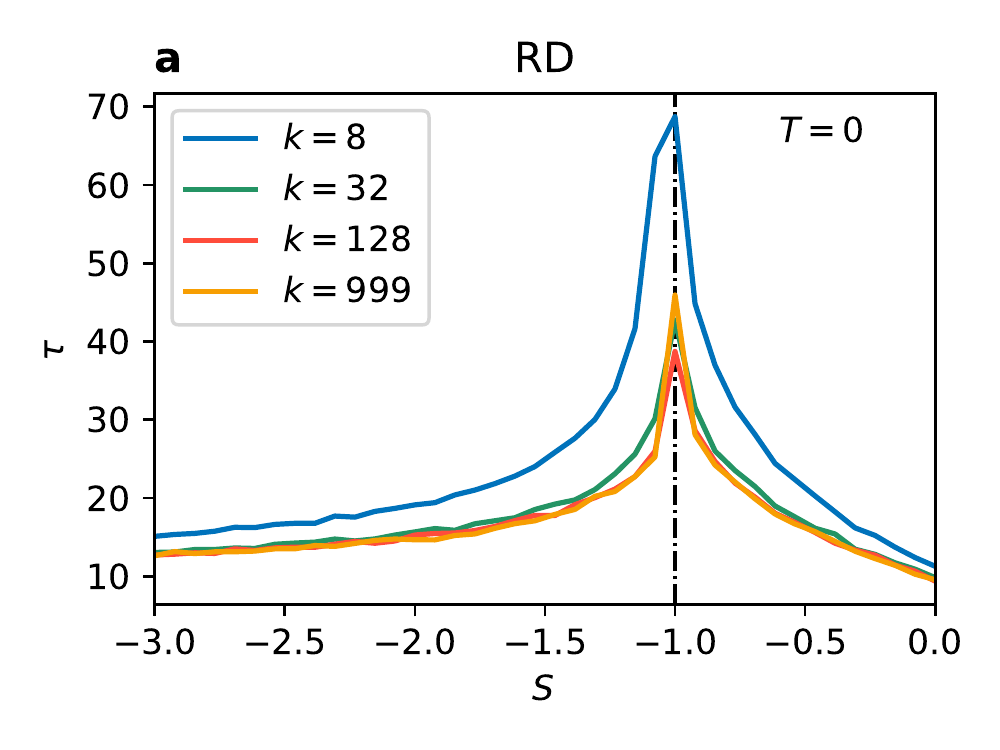}
\includegraphics[scale=0.6]{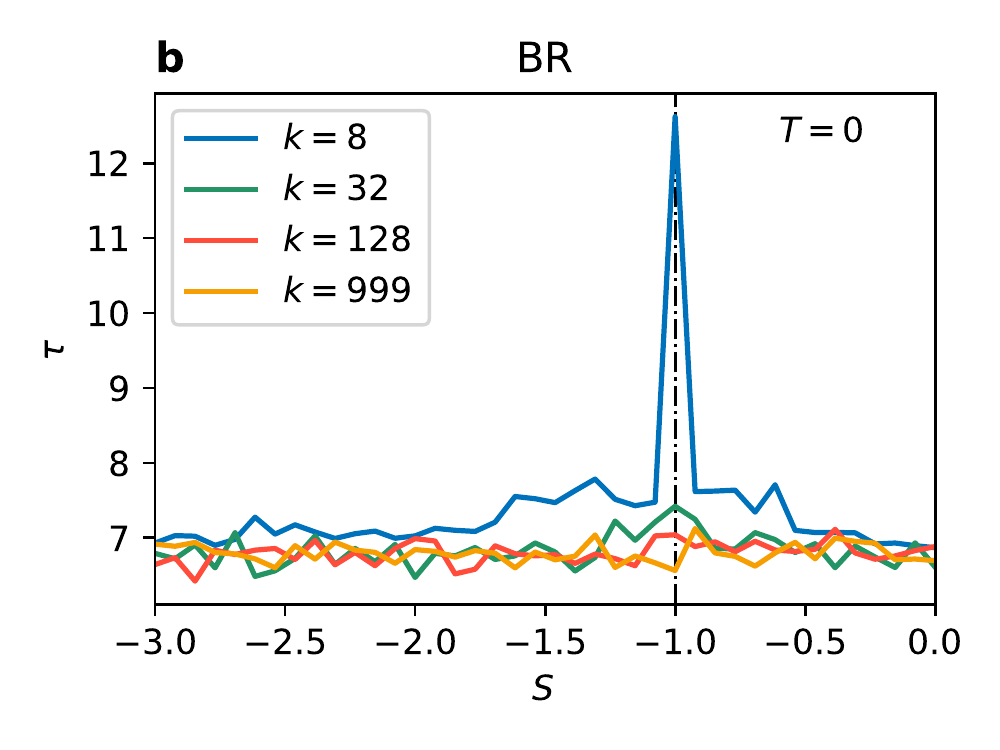}
\includegraphics[scale=0.6]{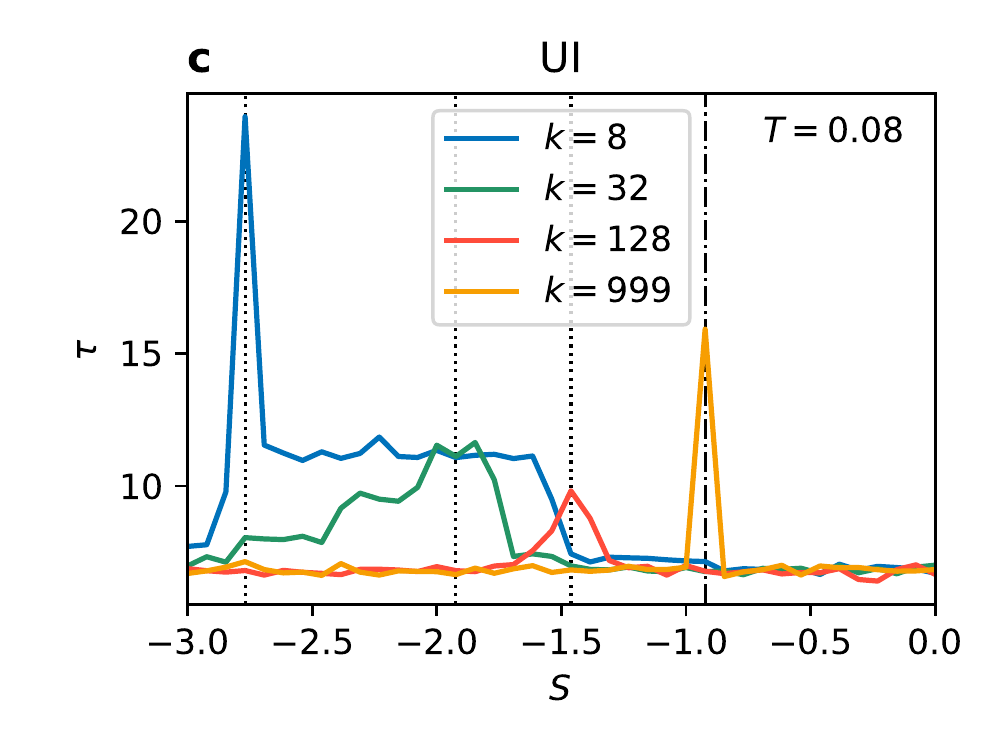}
}
\centerline{
\includegraphics[scale=0.6]{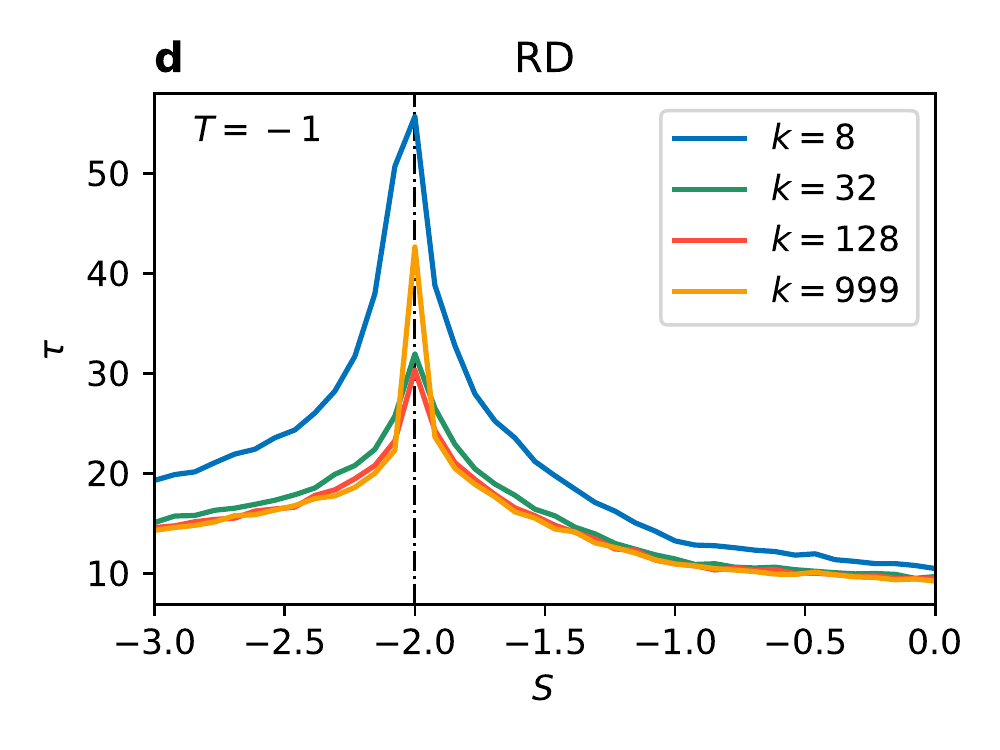}
\includegraphics[scale=0.6]{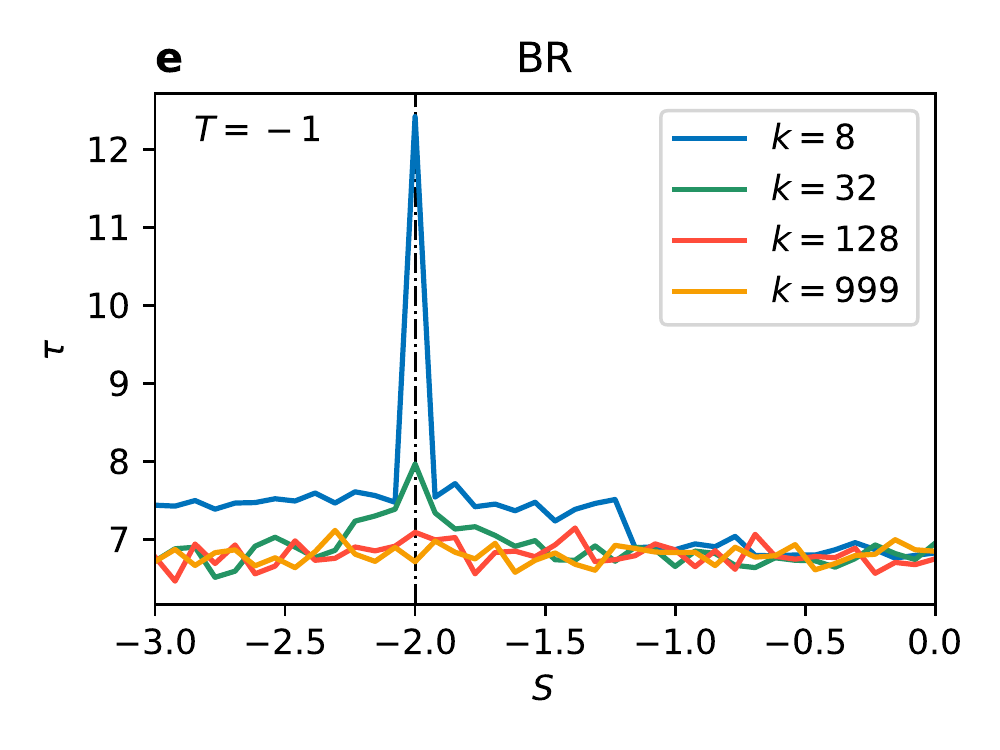}
\includegraphics[scale=0.6]{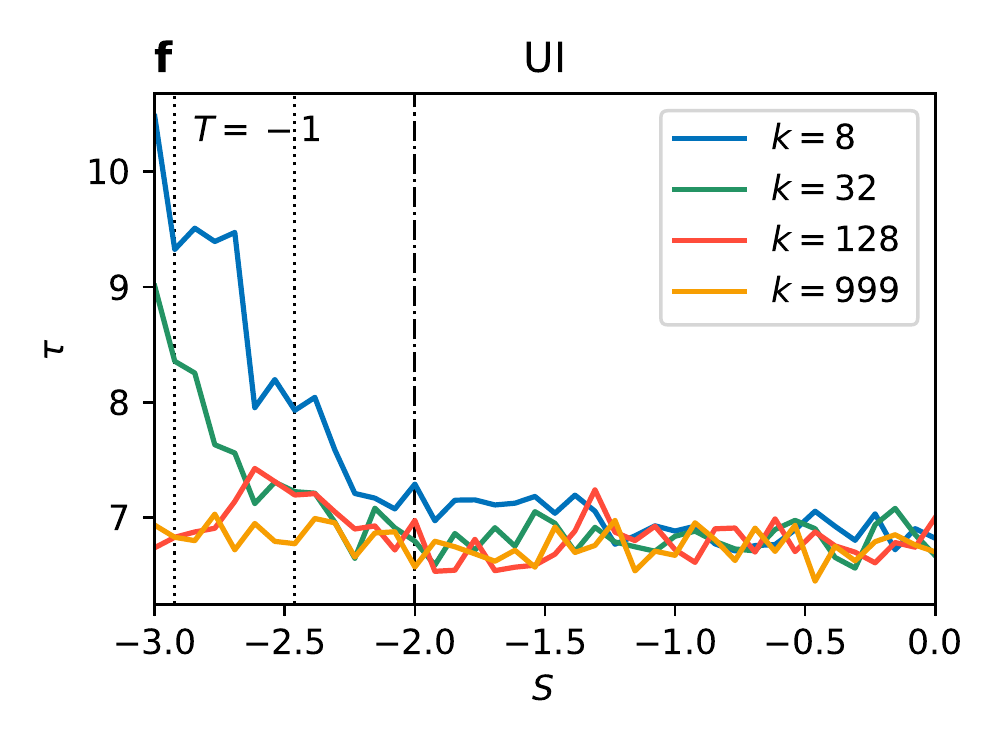}
}
\caption{Convergence time $\tau$, counted in Monte Carlo steps, in the general coordination game vs parameter
$S$ for $N=10^3$ and different values of the degree $k$.
The update rule is (a, d) the replicator dynamics, (b, e)
the best response, and (c, f) the unconditional imitation.
The upper row (a, b) presents results for $T=0$, except (c) where $T=0.08$.
The bottom row (d, e, f) presents results for $T=-1$.
Vertical dashed lines mark the value $S_c$ at which the risk-dominant strategy changes
from A to B. Dotted lines show the transition point from Figure~\ref{fig:GCG_diagram_k}
(dashed-dotted indicate both).
All values are averaged over 100 realisations.}
\label{fig:times_GCG}
\end{figure}

We next consider the average convergence time $\tau$ to the selected equilibrium.
Results for two different values of $T$ are shown in Figure~\ref{fig:times_GCG}.
As previously, RD is the update rule least affected by local effects (changes in the value of the degree).
A maximum convergence time is always found at the transition point of equilibrium selection. Additionally, the time
to reach a frozen configuration is similar
for all values of $k$, except sparse networks ($k=8$) where it becomes bigger.
When using the BR update rule converge times are in general slightly shorter.
For smaller degree ($k=8$) the transition point of equilibrium selection is firmly marked by
a narrowly peaked maximum value of $\tau$. For $k=32$ there is very small
increase in the convergence time and for larger values of degree the transition
is not identified at all. In the case of the UI update rule the transition in equilibrium selection does not coincide with the change of risk-dominant strategy and the equilibrium selection transition only manifests itself as a maximum in the convergence time for $T>0$. In the panel (c)
of Figure~\ref{fig:times_GCG} the maximum of $\tau$ is obtained at a different
point for each value of $k$, because the equilibrium selection transition moves with changing degree.
Every peak corresponds to the transition for the given $k$. For $T \leq 0$,
however, the maximum convergence time is not associated with the transition. Note, that comparing this behaviour with our results for the PCG, one must remember that in the PCG the maximum of the convergence time was associated with a transition from a frozen disorder to coordination, while in the GCG the transition is only a change of the selected equilibrium of full coordination.


\subsubsection*{The case of $T=-1$}

A particular coordination game often studied in the literature is the one described
by the following payoff matrix
\cite{eshel1998altruists,alos2006imitation,lugo2015learning,gonzalez2019coordination}:
\begin{equation}
\label{eqn:CCG_matrix}
\begin{blockarray}{ccc}
 & $B$ & $A$  \\
\begin{block}{c(cc)}
  $B$ & 1 & 0  \\
  $A$ & -b & 2  \\
\end{block}
\end{blockarray}~,
\end{equation}
where usually a restriction $b>0$ is imposed, although it is still a coordination
game up to $b>-1$. Note, that the strategy symbols changed
their positions. In a typical notation the upper left strategy is Pareto-optimal, i.e.
it is the strategy that gives the largest payoff. We denoted it before as
the strategy A. Therefore, to stay consistent
we change the denotation to maintain (A,~A) the Pareto-optimal configuration.

The game described by the payoff matrix~(\ref{eqn:CCG_matrix}) is fully
equivalent to the GCG with $T=-1$. To see this, it is enough to interchange the positions
of strategies A and B and subtract $1$ from every entry to obtain:
\begin{equation}
\label{eqn:CCG_matrix2}
\begin{blockarray}{ccc}
 & $A$ & $B$  \\
\begin{block}{c(cc)}
  $A$ & 2 & -b  \\
  $B$ & 0 & 1  \\
\end{block}
\end{blockarray}~~
\xrightarrow[\text{~~}]{\text{~subtracting 1~}}~
\begin{blockarray}{ccc}
 & $A$ & $B$  \\
\begin{block}{c(cc)}
  $A$ & 1 & -b-1  \\
  $B$ & -1 & 0  \\
\end{block}
\end{blockarray}~.
\end{equation}
In this notation it is clear that the game described in the literature
by the payoff matrix~(\ref{eqn:CCG_matrix}) is equivalent to the GCG
described by the payoff matrix~(\ref{eqn:matrix_GCG}) for $T=-1$ and $S=-b-1$.
In this parametrisation the change of risk-dominant strategy from A to B lays at $S=-2$ ($b=1$).
For $S>-2$ ($b<1$) the strategy A is risk-dominant and for $S<-2$ ($b>1$)
the strategy B is risk-dominant. For the RD or BR update rules
the transition in equilibrium selection obtained in numerical
simulations presented in Figure~\ref{fig:GCG_diagram_k}
occurs at this same point $S_c=-2$ ($b_c=1$)
(also in ER random networks, not presented due to identical character).
For this reason, here we focus on transition for the unconditional
imitation update rule

\begin{figure}[ht!]
\centerline{
\includegraphics[scale=0.65]{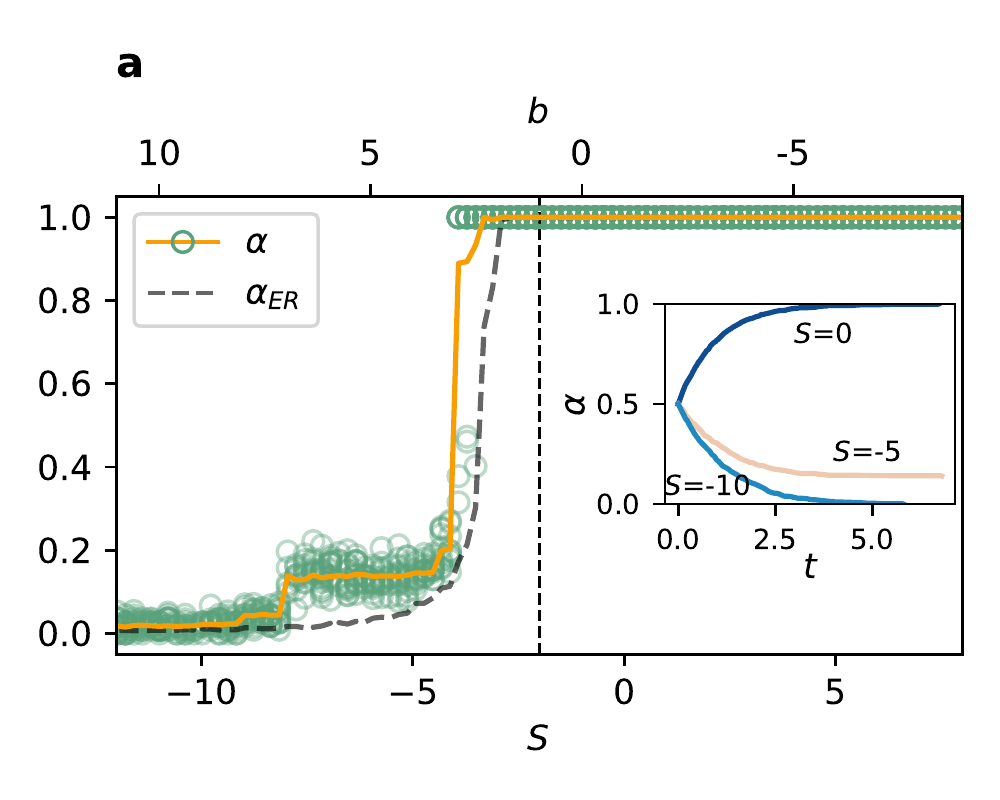}
\includegraphics[scale=0.65]{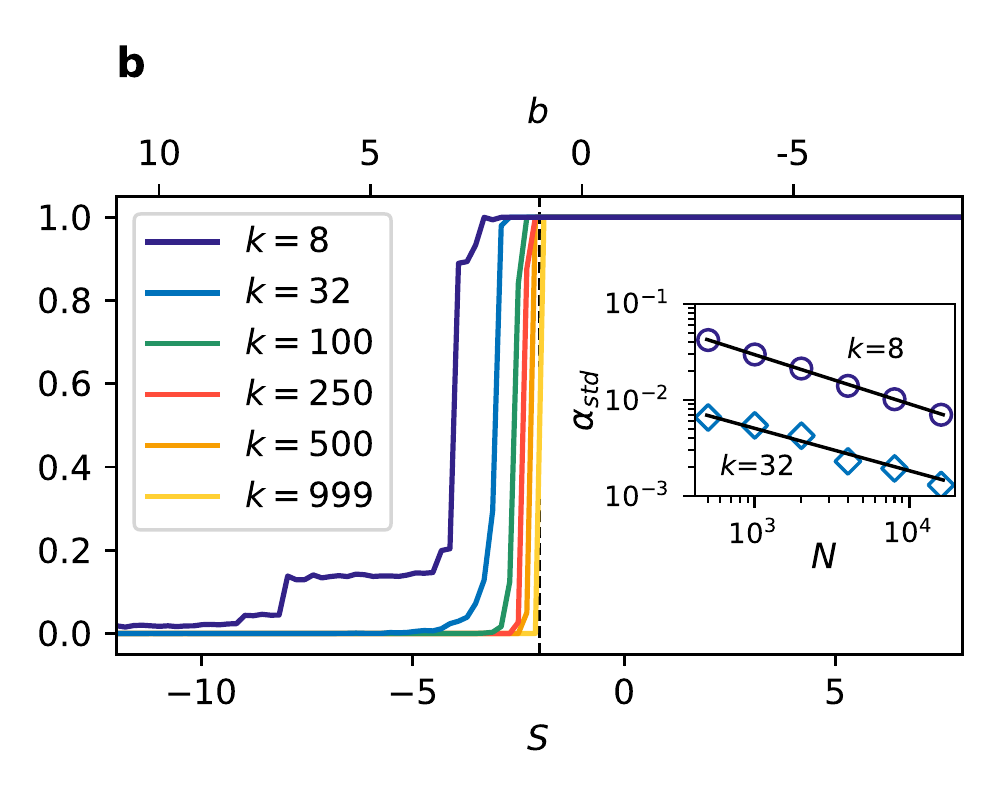}
\includegraphics[scale=0.65]{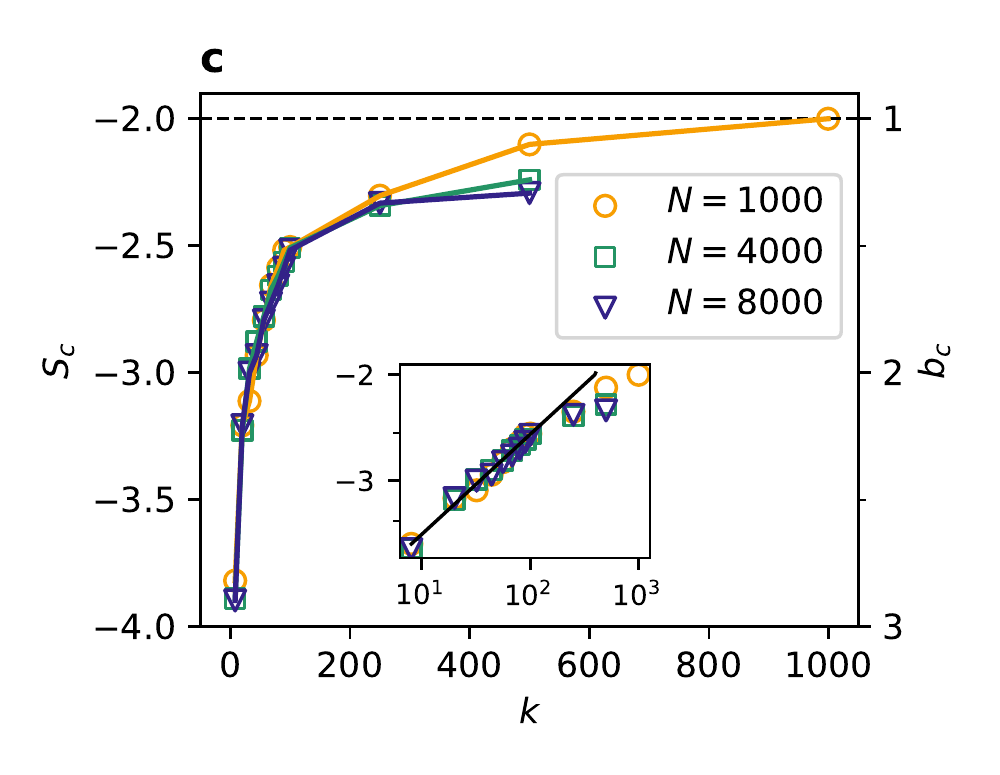}
}
\caption{
(a) Coordination rate $\alpha$ and interface density $\rho$ vs parameter $S$ (and $b$)
for $N=1000$ and $k=8$.
Each green circle represents one of 100 realisations for each value of $S$ and 
the average value is plotted with a solid line. Results are compared to the ER
random network ($\alpha_{ER}$). Inset plot: time evolution of  $\alpha$
in representative realizations for 3 values of $S$.
(b) The average value of $\alpha$ vs. parameter $S$ (and~$b$) for different
values of the degree $k$ and $N=1000$. Inset plot: scaling of the standard deviation
$\alpha_{std}$ with the network size for $S=-6$ and $k=8,k=32$;
black lines show the best power law fit.
(c) The point of transition in equilibrium selection $S_c$ (and $b_c$) vs degree $k$ for different network sizes.
Inset plot: the same results on a log-log scale with a power law fit up to $k=250$.
In all plots vertical (a, b) and horizontal (c) dashed lines mark the  point of change in risk-dominant strategy $S=-2$ ($b=1$). The update rule is unconditional imitation.
}
\label{fig:CCG_trans}
\end{figure}

In Figure~\ref{fig:CCG_trans}~(a) we present the dependence of coordination
rate $\alpha$ on the parameter $S$ (and $b$). We can see that the transition in equilibrium selection from $\alpha=0$ to $\alpha=1$
is shifted towards lower values of $S$ (or larger values of $b$) in
comparison to the values at which risk-dominant strategy changes. Interestingly, all realisations
lay very close to the average value of $\alpha$. Therefore, the coordination
is well predefined by the parameter $S$ (with other parameters fixed).
We can see that in ER random graphs the transition in equilibrium selection is slightly closer
to the $S=-2$ ($b=1$) point and also the increase of coordination rate is
smoother.
The inset of panel (a) shows representative trajectories, which quickly converge into a
frozen configuration for different parameter values.
In Figure~\ref{fig:CCG_trans}~(b) we show the average coordination rate $\alpha$
for different degrees of the network. The transition point $S_c$ of equilibrium selection from B-coordination to A-coordination shifts towards
the value $S=-2$ with growing connectivity to finally
coincide with this point of change in the risk-dominant strategy for a complete graph.
Additionally, the coordination rate changes directly from $\alpha=0$
to $\alpha=1$ for higher degree values, without the intermediate plateau 
visible for $k=8$. These results are robust regardless of the network size,
as the standard deviation $\alpha_{std}$ decreases with growing $N$
(see the inset plot of Figure~\ref{fig:CCG_trans}~(b)).
In order to investigate further the dependence of the transition point
$S_c$ (and $b_c$) on the network's degree we plot it for different
network sizes in Figure~\ref{fig:CCG_trans}~(c). Up to $k \approx 250$
the transition point follows a power law $S_c \sim k^{-0.17}$ ($R^2>0.99$)
for every size of the network. Then, the lines separate as each of them
has the upper limit of $S_c=-2$ which is obtained for a complete graph,
i.e. for $k=N-1$ which depends on the number of nodes.


\section*{Discussion}

The three update rules considered in this paper assume that a rational agent aims at increasing
its payoff. Either by directly computing the payoff and choosing the larger one
like in the BR update rule (therefore this rule requires players' knowledge about the payoff matrix), or by imitating a more successful neighbour with a larger payoff like in
the RD and UI updates rules. However, the outcome can be very different between those update rules in coordination games, even for the simplest pure coordination game.
A simple one-round two-player game is fully described by its payoff matrix, but 
an evolutionary game is defined by the update rule as much as by the payoff matrix.
We focused on the local effects and finite size effects which are crucial in networked populations. It turns out that these effects are much more important for the UI update rule, while the UI and BR updates rules are the same in a fully connected or well mixed population.

In the pure coordination game we addressed the question of when the system reaches full coordination and when it is trapped in a frozen disordered state of coexistence of the two equivalent strategies.  We found a transition from a disordered state to a state of full coordination for a critical value $k_c$ of the degree: global coordination requires a minimum connectivity of the network. The critical value is well identified by a maximal value of the convergence time. The transition is discontinuous or  first-order like for the RD and BR update rules, and continuous for the UI update rule. The value of $k_c$ is different for the different update rules, but it is system size independent for the RD and BR and system size dependent for the UI. Still, also for the UI update rule the transition remains well defined in the limit of an infinite network (thermodynamic limit), so that global coordination is obtained for a connectivity lower than
in a complete graph.

For the general coordination game we addressed the question of equilibrium selection: coordination in either a payoff-dominant or risk-dominant equilibrium. A mean field approximation for the BR or UI and the replicator equation for the RD update rule predicts that the risk-dominant equilibrium is always selected regardless if it's Pareto-optimal or not. This prediction implies that there is a transition from selecting strategy A to selecting strategy B at the point at which the risk-dominant strategy changes from A to B. This prediction is corroborated by our numerical simulations for RD and BR update rules independently of local effects (connectivity $k$) and system size. However, for the UI update rule we find that it is possible to select the Pareto-optimal strategy A even when it is not the risk-dominant one. This is a consequence of local effects -- such selection is possible only when the network is not densely connected. Our detailed analysis identifies a critical value of the parameters for this transition in equilibrium selection that depends on the network degree, or alternatively a critical value of the degree such that for lower values of the degree the Pareto-optimal strategy is selected. We note that this is an opposite effect to the one found in a circular topology \cite{alos2006imitation} where higher connectivity (but far from the complete graph limit) favours selection of the payoff-dominant strategy. In summary, it is a combination of local effects and update rule that
makes coordination on the payoff-dominant strategy A when it is not risk-dominant
possible.

Our results are general and cover many applications of
evolutionary coordination games. We shed light on the conditions
required to obtain any coordination in networked populations.
Additionally, we show when the Pareto-optimal strategy is chosen
for the coordinated state. Our work is relevant in human and animal
cooperation issues, biological, social and economical sciences, and other
fields applying evolutionary game theory. The payoff matrices we analysed
are comprehensive and cover all symmetrical coordination games
with either equivalent equilibria or with one strategy being
risk-dominant and one payoff-dominant (it can be the same strategy).
Future work could cover asymmetrical coordination games
\cite{broere2017network,bojanowski2011coordination}
such as battle of sexes and include noise or error in strategy selection,
which could further enhance coordination
\cite{roca2009imperfect,xia2012role}.


\section*{Methods}

\subsection*{Mean-field description of BR and UI}

To study the dynamics of the system we consider the coordination
rate $\alpha$ accounting for the number of nodes playing the strategy A
(divided by the network size $N$ for normalisation).
The probability of finding neighbours having a particular state
is not necessarily uniform, however the simulations suggest that strategies
are well mixed in most of the scenarios.
Therefore, we can make the mean field assumption that $\alpha k$ neighbours of a randomly
chosen node will play the strategy A and $(1-\alpha) k$ the strategy B.
Accordingly, the expected payoffs from the strategies in the general
coordination game described by the payoff matrix~(\ref{eqn:matrix_GCG})
will be:
\begin{equation}
\begin{aligned}
  &  \Pi_{\mathrm{A}} = {\alpha}  k \cdot 1 +  (1-\alpha)  k  \cdot S = k(\alpha -\alpha S +S) ,\\
  &  \Pi_{\mathrm{B}} = {\alpha}  k \cdot  T +  (1-\alpha)  k \cdot 0 = {\alpha}  k  T.
\end{aligned}
\end{equation}
When using the best response update rule every node deliberately chooses
the strategy that will result in the highest payoff,
hence the condition for the strategy A to
be chosen is simply $\Pi_{\mathrm{A}} > \Pi_{\mathrm{B}}$,
which leads to the condition:
\begin{equation}
     k(\alpha -\alpha S +S) > {\alpha}  k  T ~\implies~ 
     \alpha + \frac{S}{1-S-T} > 0,
     \label{eqn:BR_condition}
\end{equation}
with the constraints of the coordination game $S<0$ and $T<1$.
Finally, the rate of adoption of a given strategy is proportional
to the fraction of nodes using the opposite strategy leading to a time dependence described by:
\begin{equation}
    \frac{\partial {\alpha}}{\partial t} = 
    \frac{k}{N} \left[ \theta \left({\alpha} + \frac{S}{1-S-T} \right) - {\alpha} \right],
    \label{eqn:mean_field_BR_UI}
\end{equation}
where $\theta$ is the Heaviside step function. If we take $\alpha = 0.5$,
which is the initial value used in the simulations,
in the equation above we can see which strategy is selected.
More precisely, from the argument of the Heaviside function we obtain the
inequality~(\ref{eqn:BR_condition}), but with $\alpha = 0.5$, from which
we obtain a condition for strategies to be evolutionary chosen:
$T<S+1$ for the strategy A and the opposite for
the strategy B. Note, that this is the same condition as for strategies
to be risk-dominant, so that the mean field approximation predicts that the risk-dominant equilibrium is selected. The same analysis can be applied
to unconditional imitation
with large average degree. The active node will imitate the most
successful neighbour in the network, which in a complete graph is
the best choice in terms of the future payoff, because all nodes
have the same neighbourhood. Therefore, effectively
the active node will choose the most lucrative strategy as
in the best response. This, however,
might not be true in sparse networks. 

\subsection*{Replicator equation for RD}

A population of players using replicator dynamics
is described by the general replicator equation:
\begin{equation}
    \dot{x_i} = x_i ( \hat{e}_i \cdot M \vec{x} - \vec{x} \cdot M \vec{x} ) ,
\end{equation}
where $M$ is the payoff matrix and $\vec{x} = [x_A , x_B]$ accounts for
the fraction $x_A$ of individuals using the 
strategy A and $x_B$ using the strategy B,
i.e.  $x_A=\alpha$ and $x_B=1-\alpha$.
Therefore, $\hat{e}_i \cdot M \vec{x}$
is the average payoff of individuals playing strategy $i$ and 
$\vec{x} \cdot M \vec{x}$ is the average payoff in the whole population. Using
the constraint $x_B = 1 - x_A$ and the exact form of the payoff matrix $M$
in the general coordination game~(\ref{eqn:matrix_GCG})
we obtain the equation for dynamics of the coordination
rate $\alpha=x_A$:
\begin{equation}
   \frac{\partial {\alpha}}{\partial t} = (S+T-1) \alpha^3
   + (1-2S-T) \alpha^2 +S \alpha .
    \label{eqn:replicator_eqn}
\end{equation}
It has three stationary solutions at
$\alpha^* \in \{ 0,  \frac{S}{S+T-1}, 1 \}$. However, the
solution with no full coordination $\alpha^*=  \frac{S}{S+T-1}$ is linearly unstable because the inequality
$S+T-1 <0$ is always satisfied for a coordination game ($S<0$ and $T<1$).
The change of sign of $ \frac{\partial {\alpha}}{\partial t}$ in (\ref{eqn:replicator_eqn}) also takes place at
$\alpha^*= \frac{S}{S+T-1}$. Therefore, in order to reach the coordination
at the strategy~A ($\alpha=1$) the condition:
\begin{equation}
     \alpha > \frac{S}{S+T-1} 
\end{equation}
must be fulfilled, which is equivalent to the condition~(\ref{eqn:BR_condition})
obtained in the mean-field approach for the BR. Taking into account the
initial condition of $\alpha =0.5$ we obtain the inequality
$T<S+1$ for the strategy A to be evolutionary chosen. Therefore, the RD equation also predicts that that the risk-dominant equilibrium is selected.

\begin{figure}[ht]
\centering
\includegraphics[scale=0.5]{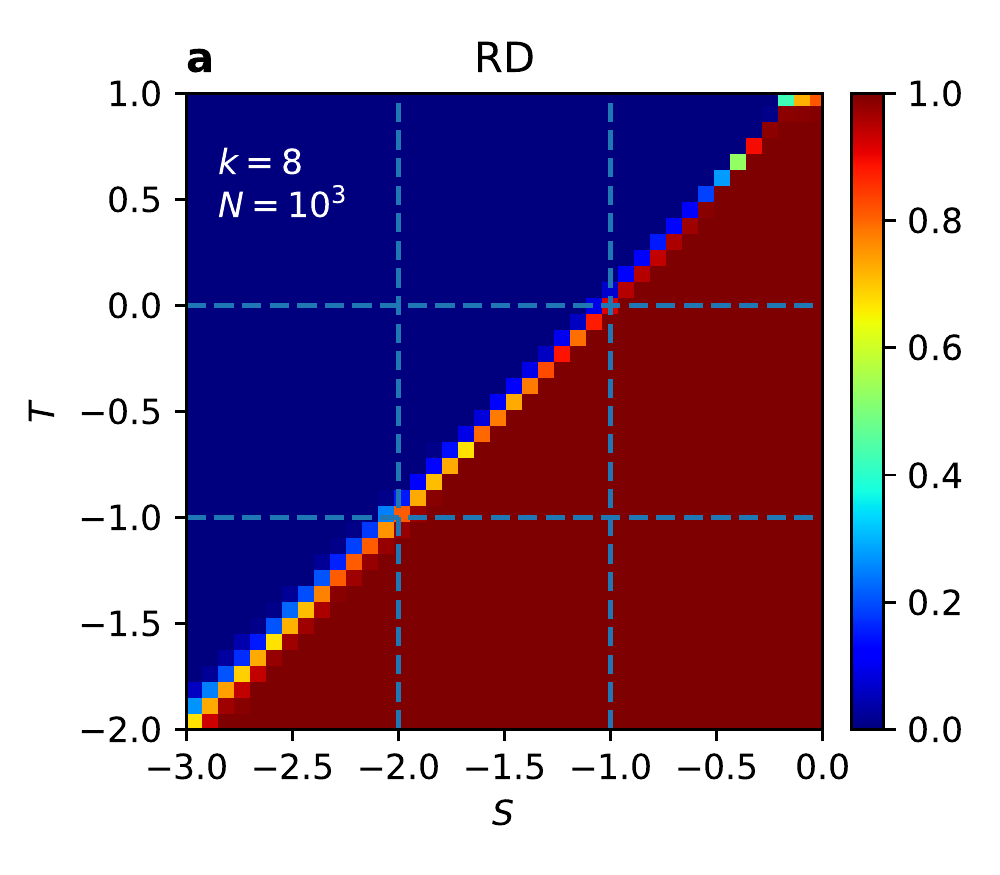}
\includegraphics[scale=0.5]{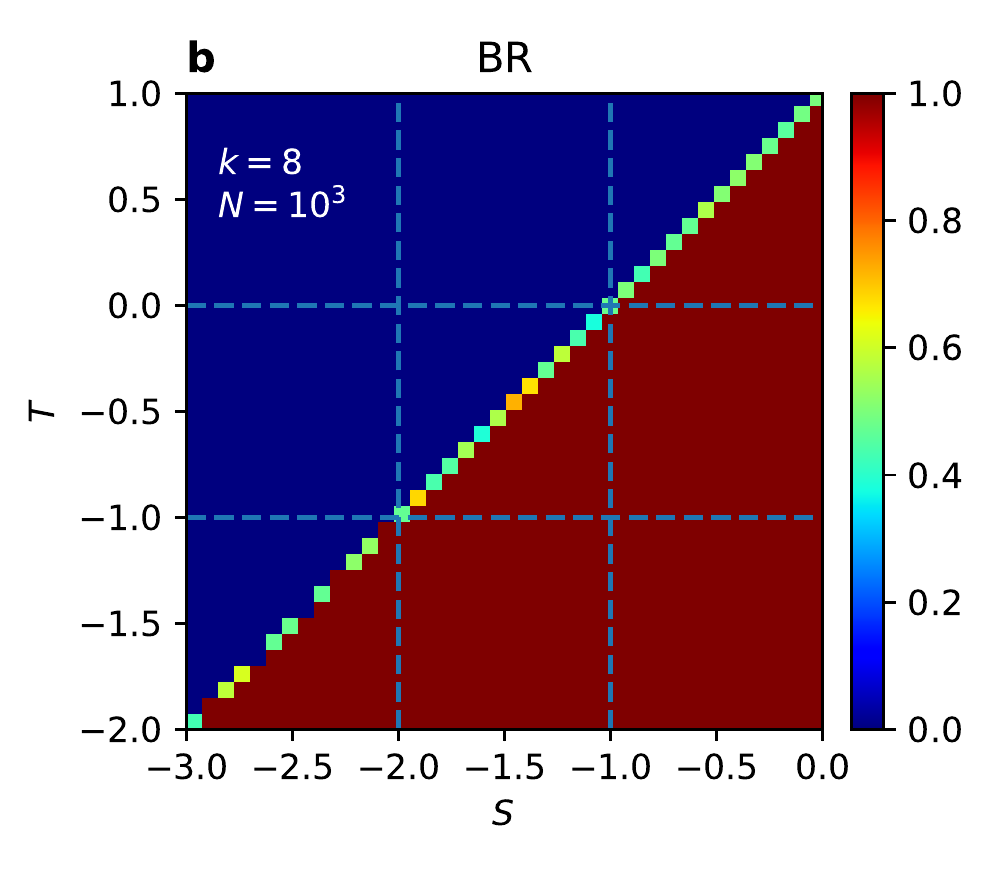}
\includegraphics[scale=0.5]{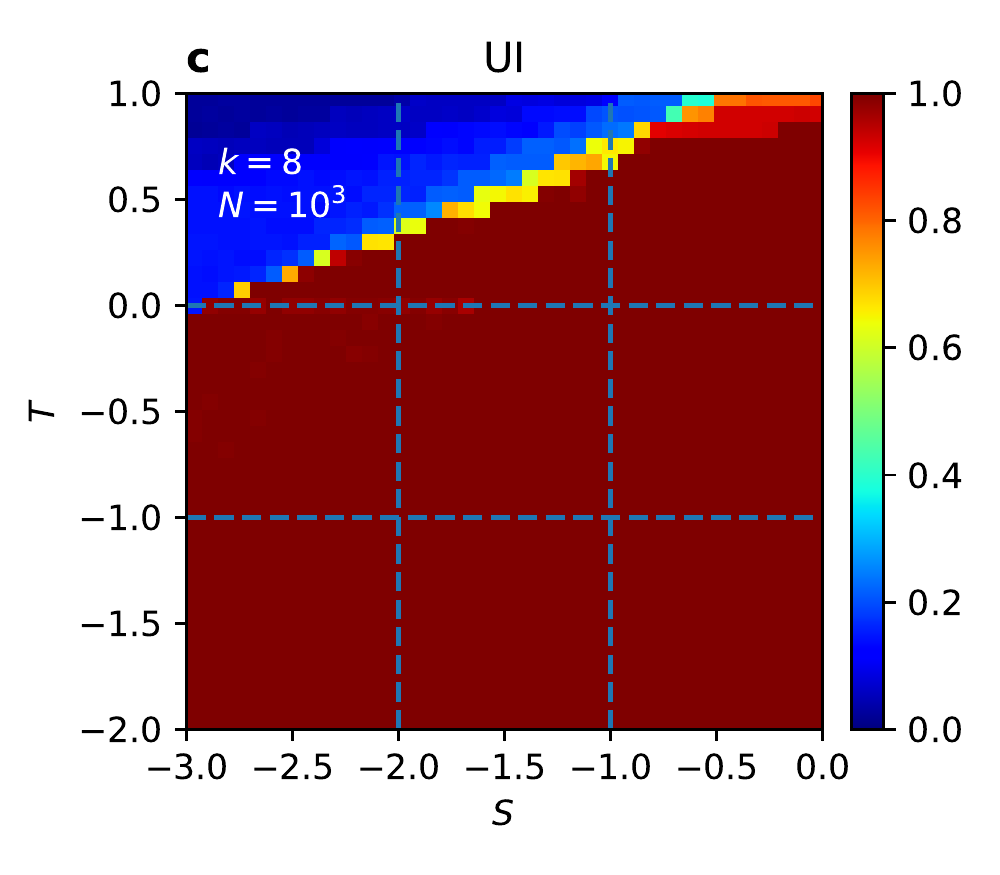}
\includegraphics[scale=0.5]{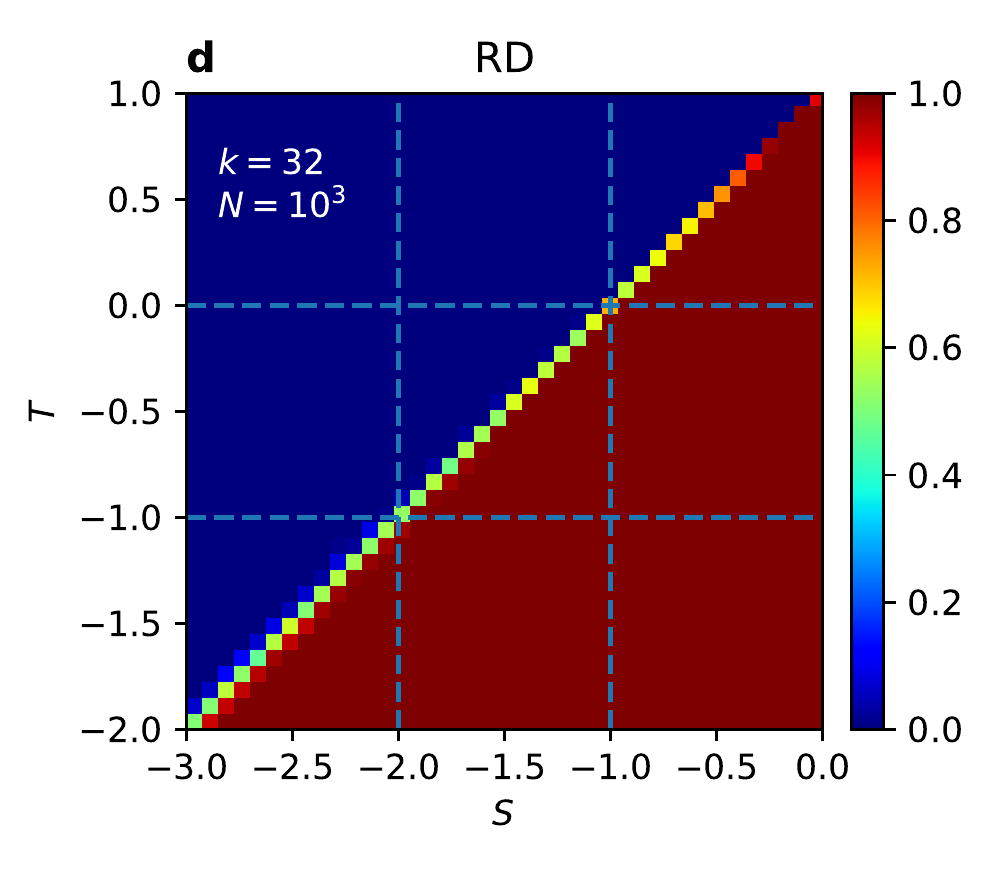}
\includegraphics[scale=0.5]{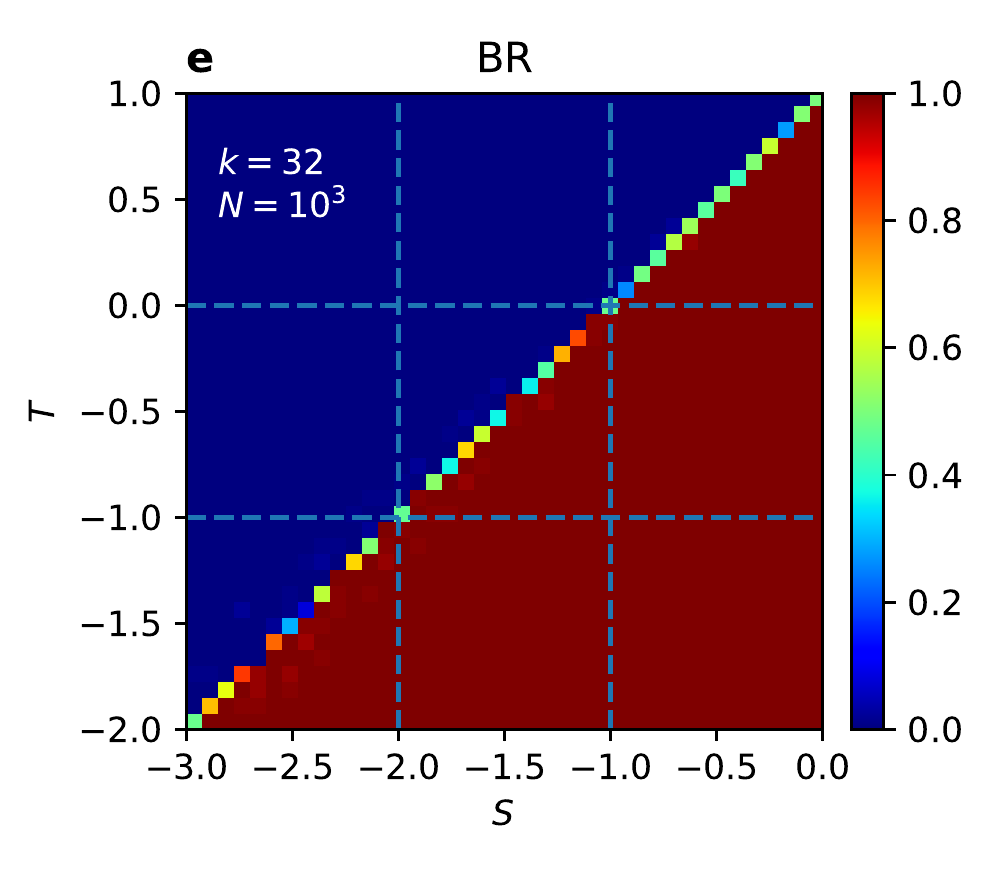}
\includegraphics[scale=0.5]{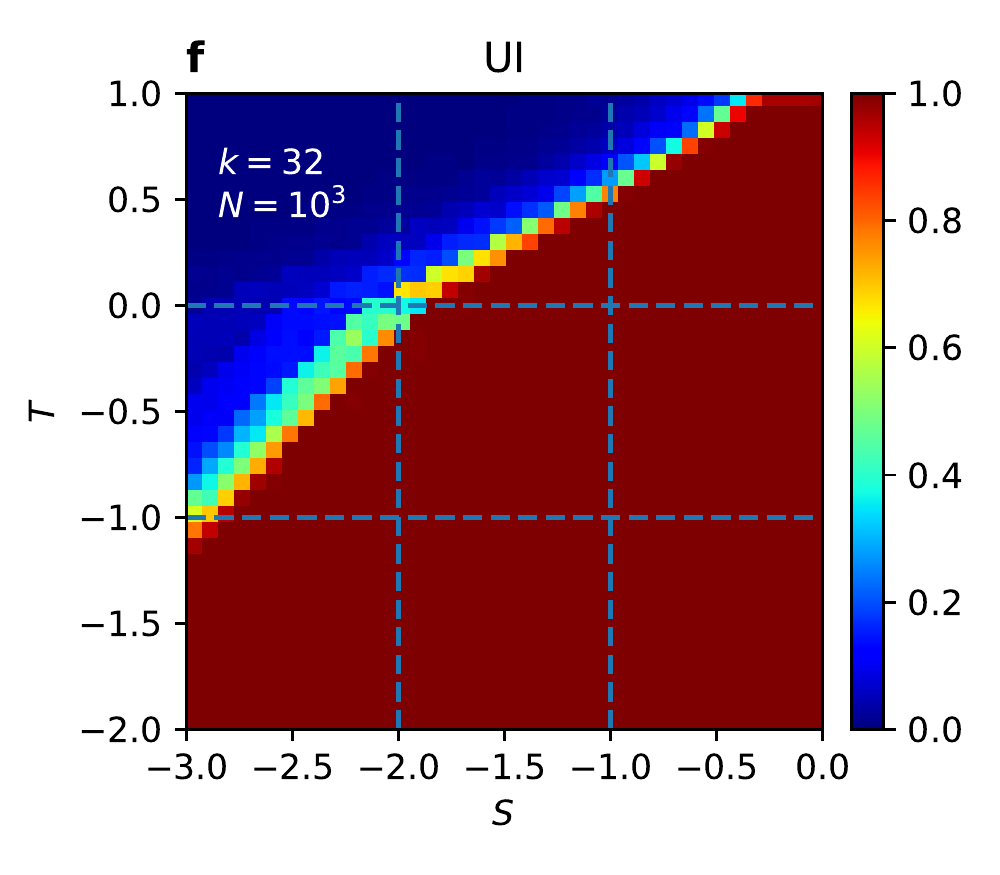}
\includegraphics[scale=0.5]{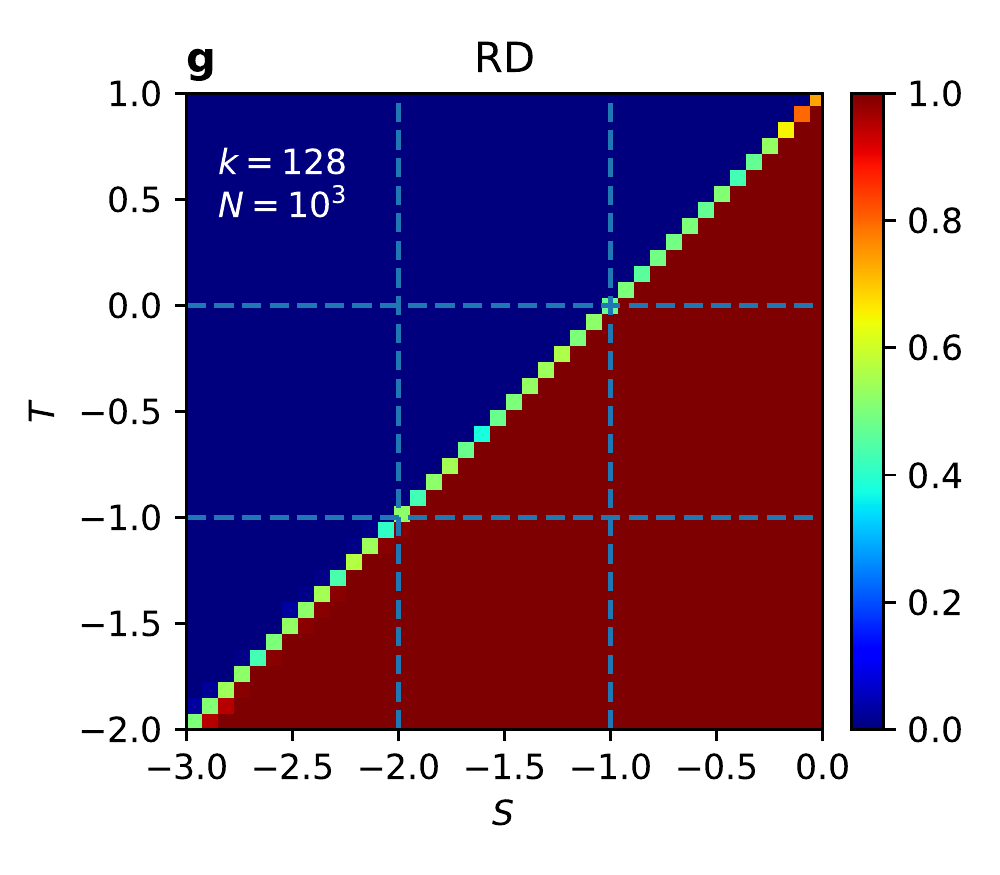}
\includegraphics[scale=0.5]{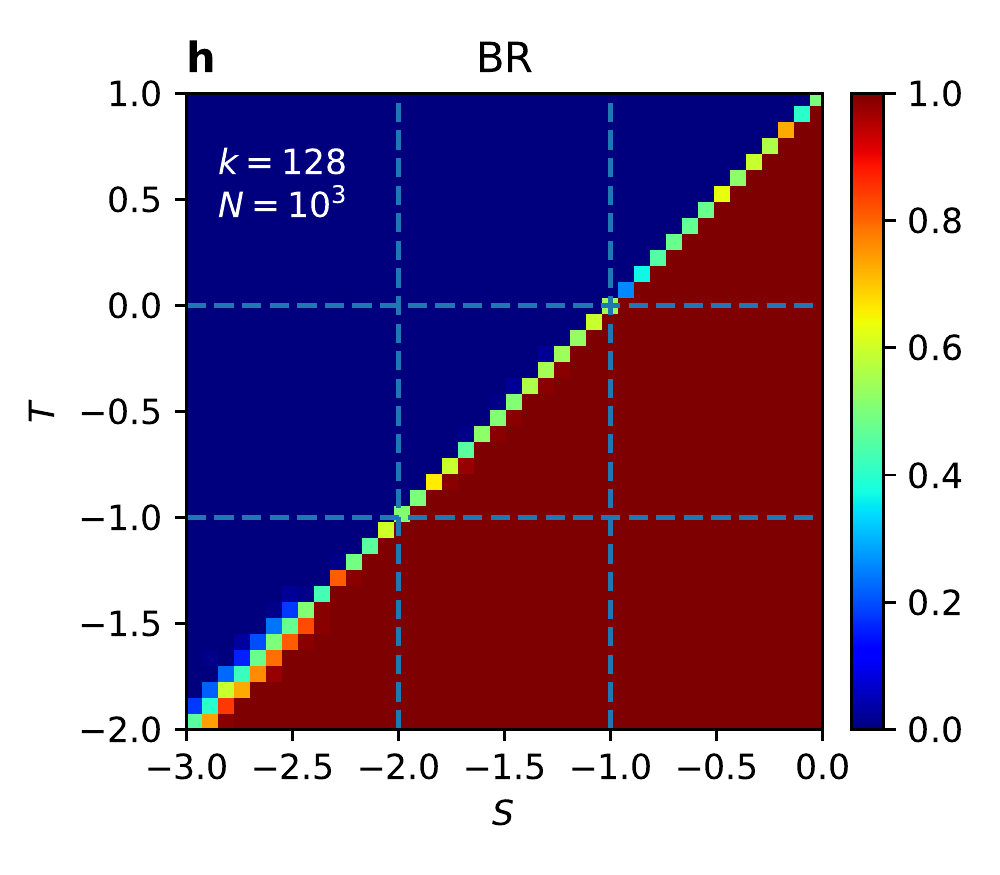}
\includegraphics[scale=0.5]{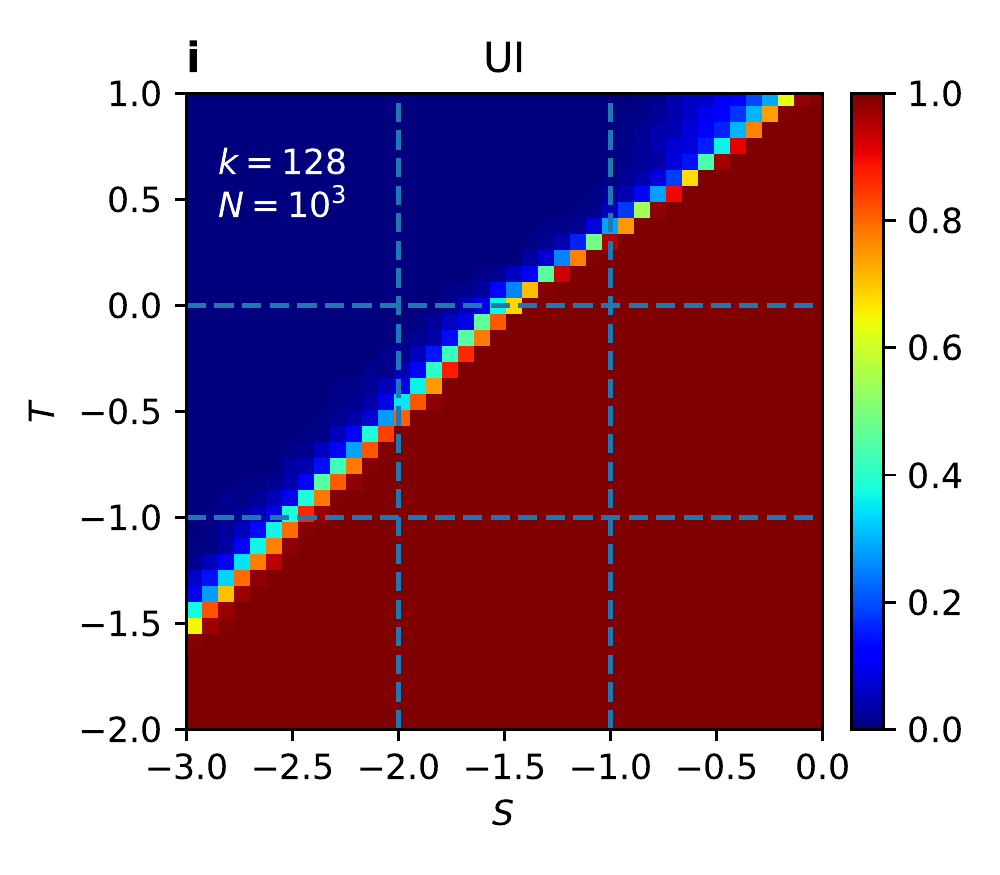}
\includegraphics[scale=0.5]{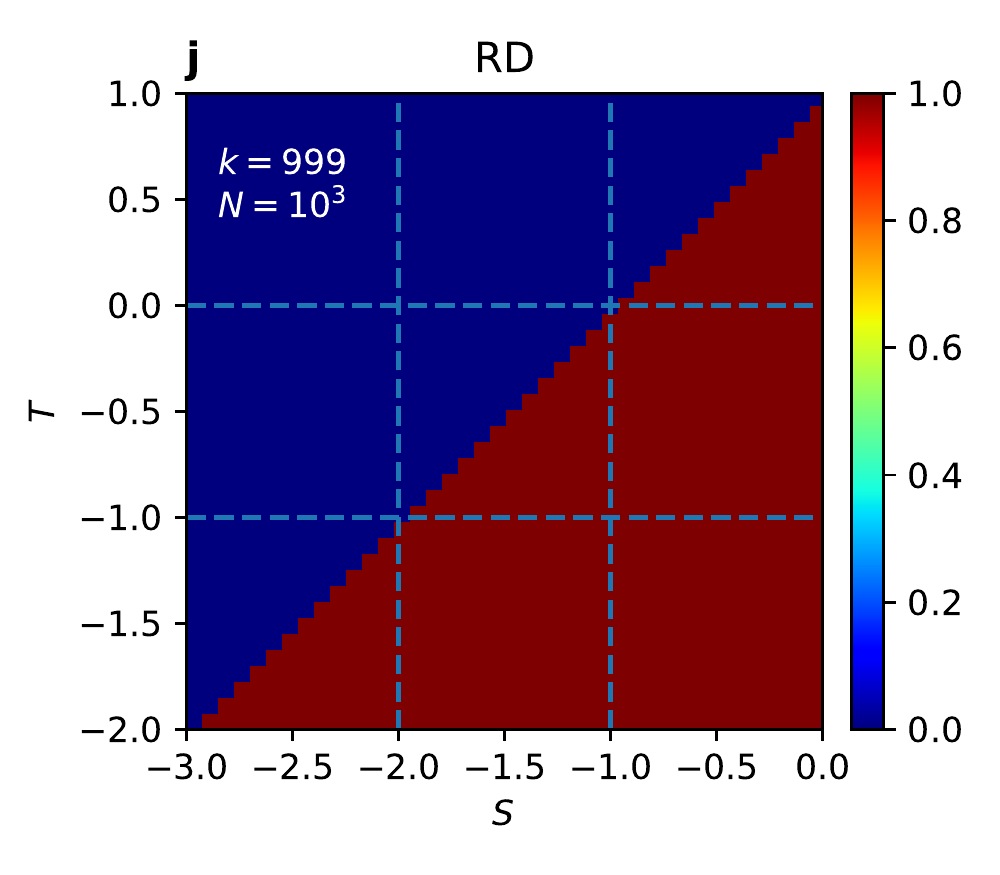}
\includegraphics[scale=0.5]{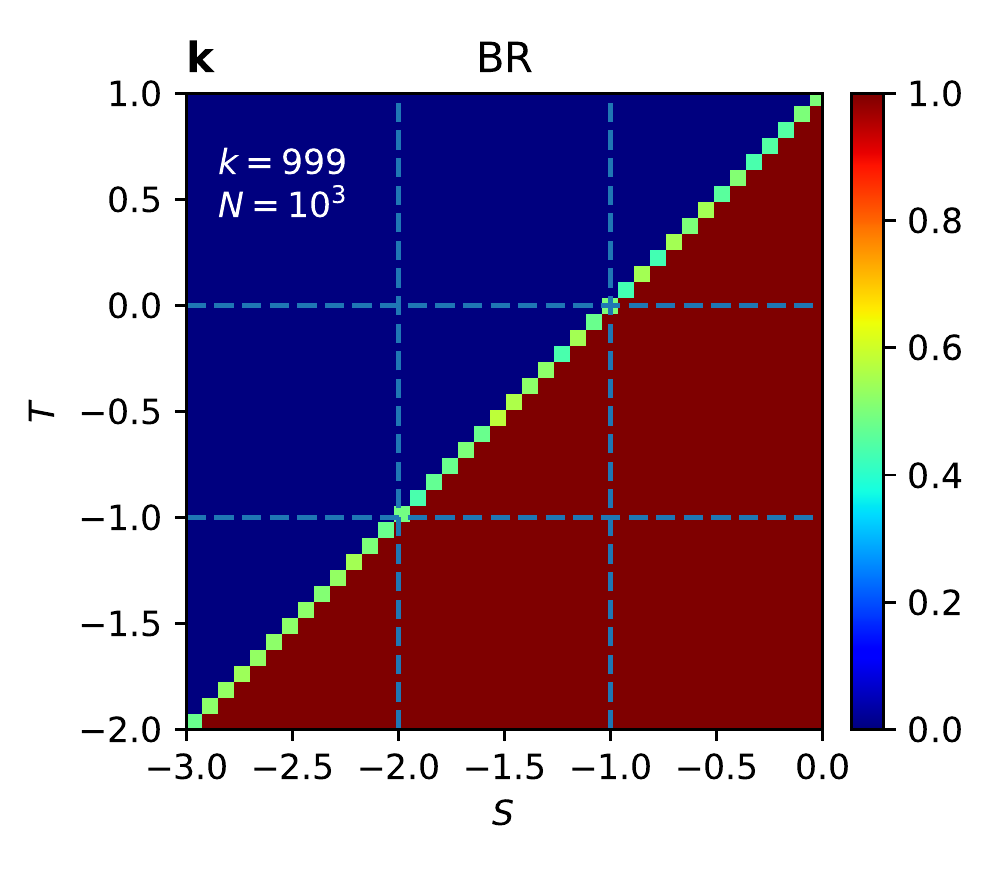}
\includegraphics[scale=0.5]{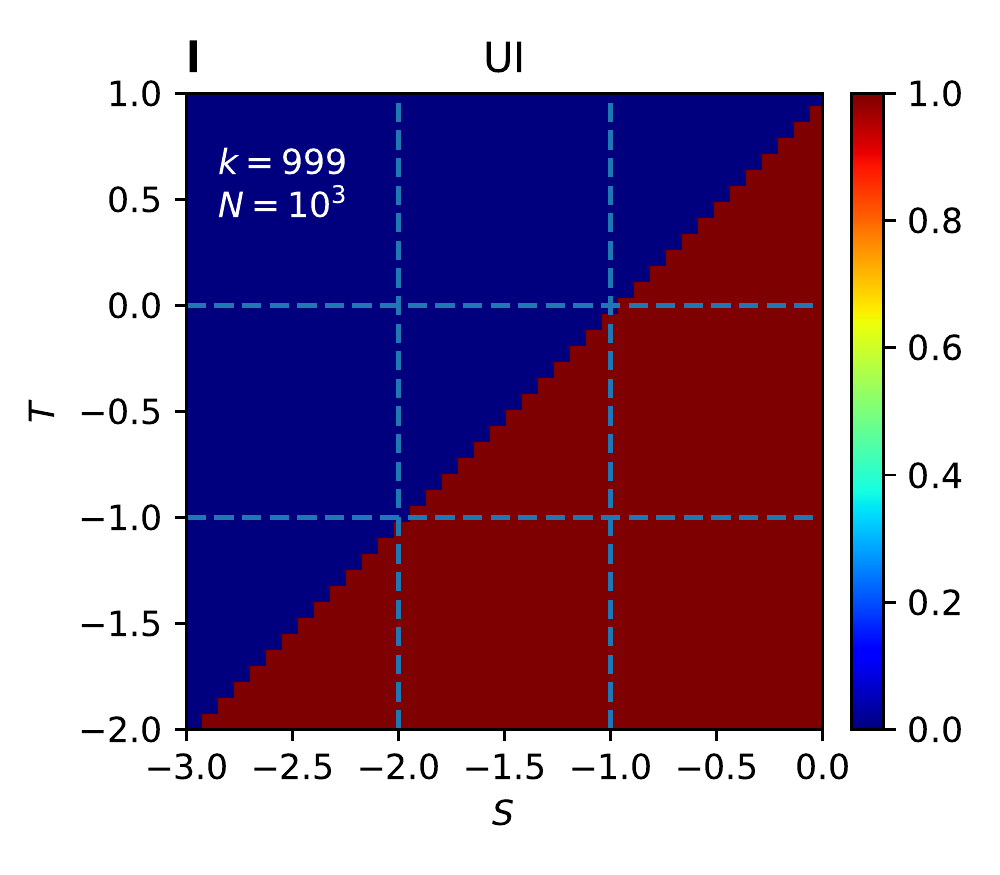}
\includegraphics[scale=0.5]{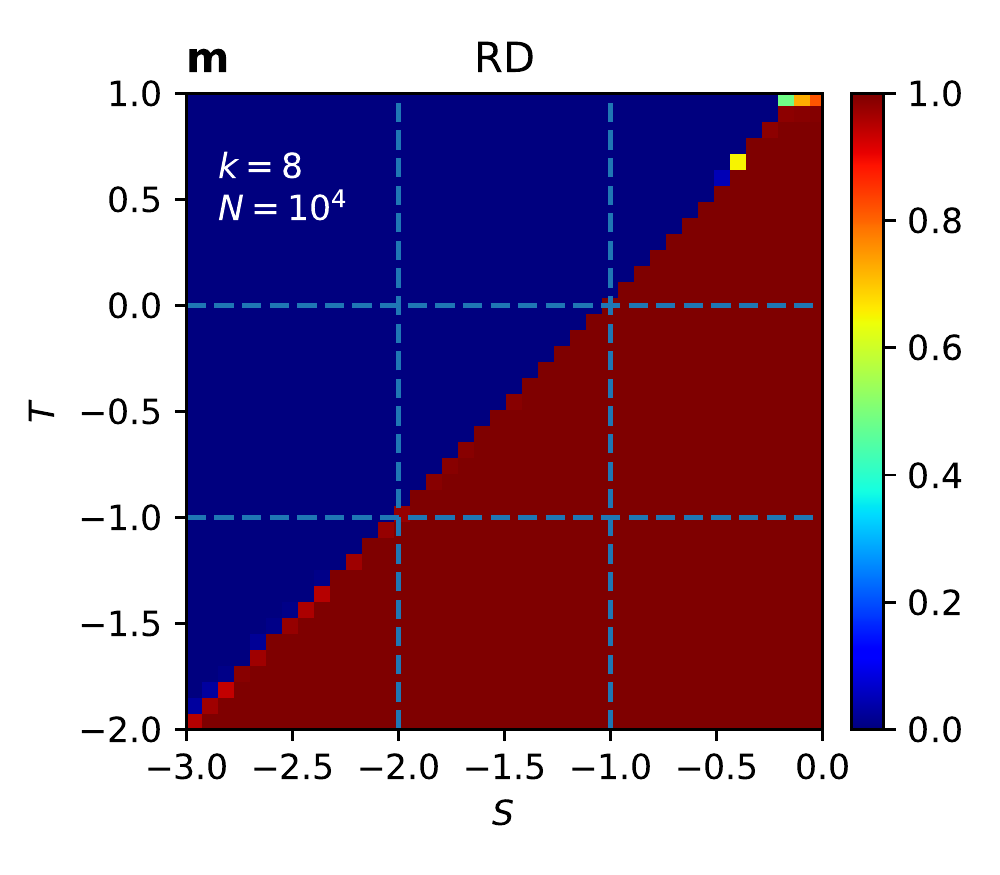}
\includegraphics[scale=0.5]{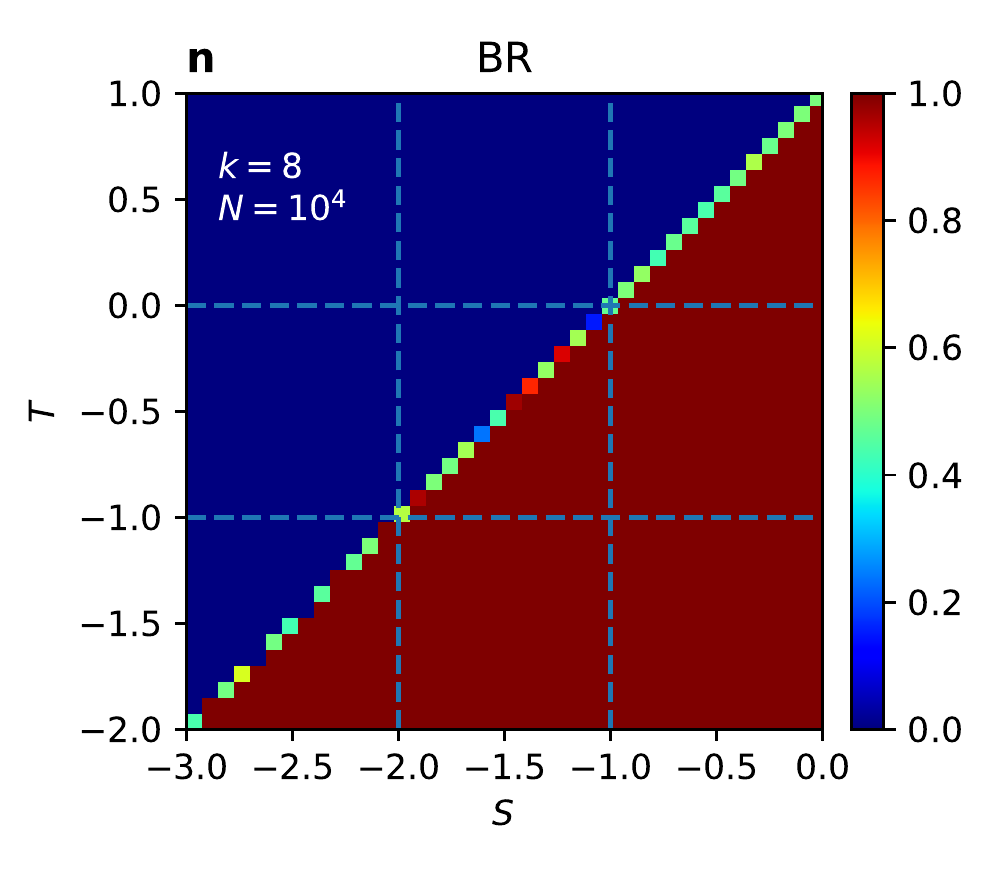}
\includegraphics[scale=0.5]{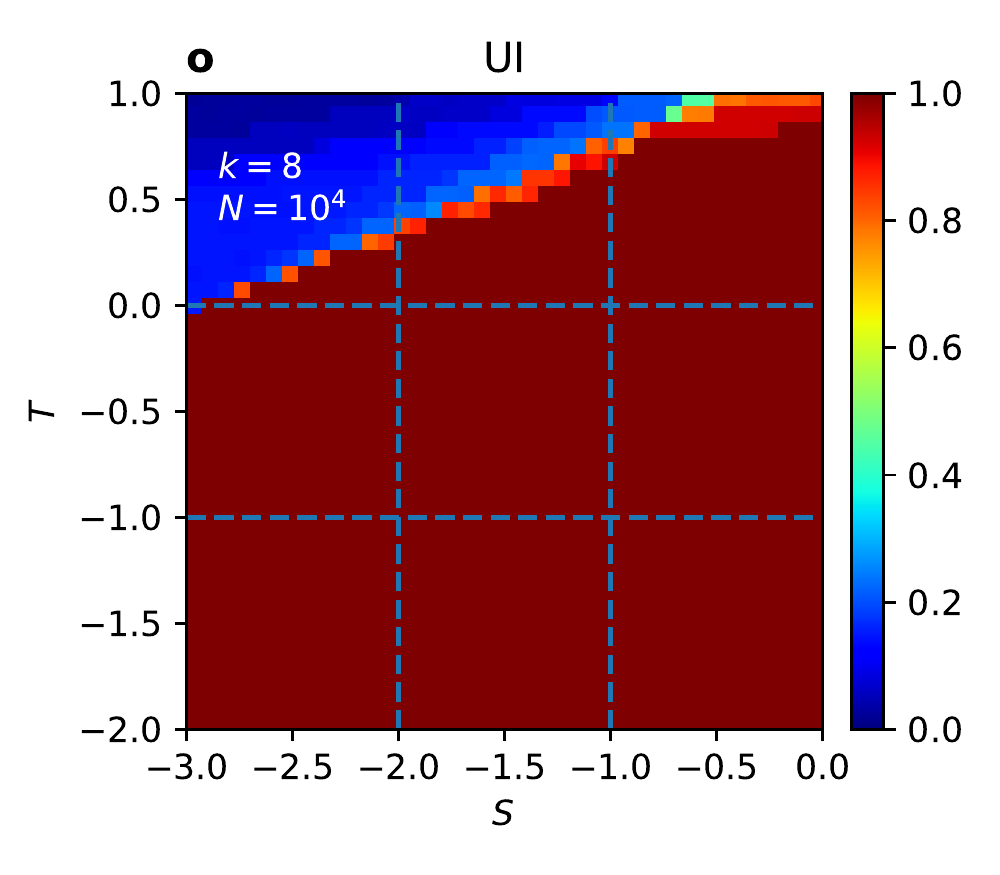}
\caption{
Phase diagram of the general coordination game --
coordination rate $\alpha$ vs parameters $S$ and $T$
for (a, d, g, j, m) the replicator dynamics,
(b, e, h, k, n) the best response, and (c, f, i, l, o) the unconditional imitation
update rule.
The diagrams are computed for $N=10^3$ and different values of degree $k$:
(a, b, c) $k=8$, (d, e, f) $k=32$, (g, h, i) $k=128$, and (j, k, l)
$k=999$ (complete graph), except the bottom row (m, n, o) which presents diagrams
for $N=10^4$ and $k=8$.
Results are obtained from 100 realisations.
}
\label{fig:GCG_coord_k}
\end{figure}


\section*{Acknowledgements}

Financial support has been received from the Agencia Estatal de Investigacion
(AEI, MCI, Spain) and Fondo Europeo de Desarrollo Regional (FEDER, UE),
under Project PACSS (RTI2018-093732-B-C21/C22) and the Maria de Maeztu Program
for units of Excellence in R\&D (MDM-2017-0711). 
T.R. would like to acknowledge the support from the Polish National Science Centre
under Grant No. 2019/32/T/ST2/00133.

\section*{Author contributions}

M.S.M. and T.R. conceived and designed the research, T.R. conducted the simulations,
M.S.M. and T.R. analysed the results, wrote and reviewed the manuscript.

\section*{Competing interests}

The authors declare no competing interests.

\section*{Additional information}

\textbf{Supplementary information} is available for this paper at Figure~\ref{fig:GCG_coord_k}.

\end{document}